\documentclass[12pt]{article}
\usepackage{latexsym}
\usepackage{bbm}
\usepackage{amsopn}
\usepackage{amsfonts}
\usepackage{amssymb}
\usepackage{amsthm}
\usepackage{amsmath}

\newtheorem{Theorem}{Theorem}[section]

\newtheorem{Lemma}[Theorem]{Lemma}
\newtheorem{Corollary}[Theorem]{Corollary}

\renewcommand{\theequation}{\arabic{section}.\arabic{equation}}

\newcommand{\tree}{\mathcal{T}}

\begin{document}


\title{\bf On the Structure of the Observable Algebra of QCD on the Lattice}

\author{
    P. D. Jarvis\\
    School of Mathematics and Physics, University of Tasmania\\
    GPO Box 252-21, Hobart Tas 7001, Australia\\
    \ \\
    J. Kijowski \\
    Center for Theoretical  Physics, Polish Academy of Sciences\\
    al. Lotnik\'ow 32/46, 02-668 Warsaw, Poland\\
    \ \\
    G. Rudolph \\
    Institut f\"ur Theoretische Physik, Universit\"at Leipzig\\
    Augustusplatz 10/11, 04109 Leipzig, Germany\\
    }

\maketitle


\begin{abstract}
The structure of the observable algebra ${\mathfrak O}_{\Lambda}$
of lattice QCD in the Hamiltonian approach is investigated. As was
shown earlier, ${\mathfrak O}_{\Lambda}$ is isomorphic to the
tensor product of a gluonic $C^{*}$-subalgebra, built from gauge
fields and a hadronic subalgebra constructed from gauge invariant
combinations of quark fields. The gluonic component is isomorphic
to a standard CCR algebra over the group manifold $SU(3)$. The
structure of the hadronic part, as presented in terms of a number
of generators and relations, is studied in detail.
It is shown that its irreducible representations are classified by triality.
Using this, it is 
proved that the hadronic algebra is isomorphic to the commutant of the 
triality operator in the enveloping algebra of the Lie super algebra 
${\rm sl(1/n)}$ (factorized by a certain ideal).  
\end{abstract}

\newpage

\vspace{0.5cm}

\tableofcontents

\newpage

\setcounter{equation}{0}
\section{Introduction}

This paper is a continuation of \cite{KR} and \cite{KR1}, where we
have investigated quantum chromodynamics (QCD) on a finite lattice
in the Hamiltonian approach. In \cite{KR} we have analyzed the
structure of the field algebra of QCD and the Gauss law.
Concerning the latter, there is a remarkable difference comparing
with quantum electrodynamics (QED). In QED, we have a {\em local}
Gauss law, which is built from gauge invariant operators and which
is linear. Thus, one can ``sum up'' the local Gauss laws over all
points of a given (spacelike) hyperplane in space time yielding
the following gauge invariant conservation law: The global
electric charge is equal to the electric flux through a $2$-sphere
at infinity. In QCD the local Gauss law is neither built from
gauge invariant operators nor is it linear, but it is possible to
extract a gauge invariant, additive law for operators with
eigenvalues in the dual of the center of $SU(3)$, (which is
identified with ${\mathbb Z}_3$). This implies -- as in QED -- a
gauge invariant conservation law: The global ${\mathbb
Z}_3$-valued colour charge (triality) is equal to a ${\mathbb
Z}_3$-valued gauge invariant quantity obtained from the colour
electric flux at infinity. We stress that the notion of triality
occurred in the literature a long time ago. On the level of lattice
gauge theories, this notion is already implicitly contained in a
paper by Kogut and Susskind, see \cite{Kogut}. In particular, Mack
\cite{Mack} used it to propose a certain (heuristic) scheme of
colour screening and quark confinement, based upon a dynamical
Higgs mechanism with Higgs fields built from gluons. For similar
ideas we also refer to papers by `t Hooft, see \cite{tHooft} and
references therein. This concept was also used in a paper by Borgs
and Seiler \cite{Borgs}, where the confinement problem for
Yang-Mills theories with static quark sources at nonzero
temperature was discussed. In this context, also the Gauss law for
colour charge was analyzed.

In \cite{KR1} we have analyzed the observable algebra of QCD in
the above context. The observable algebra ${\mathfrak
O}_{\Lambda}$ is obtained by imposing gauge invariance and the
local Gauss law. It turns out that ${\mathfrak O}_{\Lambda}$ is
isomorphic to the tensor product of a gluonic $C^{*}$-subalgebra,
built from lattice gauge fields and a hadronic subalgebra
constructed from gauge invariant combinations of quark fields. The
gluonic component is isomorphic to a standard CCR algebra over the
group manifold $SU(3)$, whereas the hadronic subalgebra
${\mathfrak O}^{mat}_{\tree}$ is built from bilinear and trilinear
gauge invariant combinations of the quark fields. We show that it
is isomorphic to the commutant of the triality operator in the
CAR-algebra generated by $n = 12N$ creation and annihilation
operators. Here, $N$ is the number of lattice sites. This fact
enables us, to prove in an elegant way that irreducible
representations of ${\mathfrak O}_{\Lambda}$ are labelled by
triality. Moreover, we show that ${\mathfrak O}^{mat}_{\tree}$ is
isomorphic to the commutant of the triality operator in the
enveloping algebra of the Lie superalgebra ${\rm sl(1/n)}$,
factorized by a certain ideal. In this language, the $3$
inequivalent representations naturally arise via a standard Kac
module construction. This classification result confirms the
classification of irreducible representations of ${\mathfrak
O}_{\Lambda}$ obtained in \cite{KR1} by a completely different
method. We believe that the various presentations of ${\mathfrak
O}^{mat}_{\tree}$ found here will be crucial in future
investigations of dynamical problems of QCD in terms of
observables.

The presentation of ${\mathfrak O}_{\Lambda}$ used in this paper
is based upon a certain gauge fixing procedure, which works well
on the generic stratum of the action of the gauge group on the
underlying classical configuration space. First steps towards
including non-generic strata have been made as well \cite{ChKRS}.
It is worthwile to try to omit the gauge fixing philosophy and to
analyze ${\mathfrak O}_{\Lambda}$ in more intrinsic terms. This
leads to polynomial super algebras, see \cite{JR}.

We stress that a similar analysis has been performed for
(spinorial and scalar) QED, see \cite{KRT}, \cite{KRS} and
\cite{KRS1}. There, the matter field part of the observable
algebra is generated by an ordinary Lie algebra. For QCD, we get a
Lie superalgebra, because here we have additionally trilinear
gauge invariant operators (of baryonic type) built from matter
fields.

Finally, we note that standard methods from algebraic quantum
field theory for models which do not contain massless particles,
see \cite{DHR}, do not apply here. For an analysis of problems
with massless particles within this approach we refer to
\cite{SW1}, \cite{S}, \cite{CH}, \cite{JF}, \cite{Bu}, \cite{FM} and further
references therein. For basic notions concerning lattice gauge
theories (including fermions), see \cite{Seiler} and references
therein.

Our paper is organized as follows: to keep the paper
self-contained, in Sections \ref{Algebra}, \ref{Gauge--Gauss} and
\ref{Observablealgebra} we briefly summarize the results of our
previous papers. In Subsection \ref{gen--rel} we present a
systematic study of ${\mathfrak O}^{mat}_{\tree}$ in terms of
generators and relations. Next, in Subsection \ref{structure-j-w},
we reduce the set of relations to a certain minimal set and in
Subsection \ref{irreps-rozdzial} we present a classification of
irreducible representations of ${\mathfrak O}^{mat}_{\tree}$.
Finally, in Subsection \ref{Super Lie Structure}, the above
mentioned super Lie structure is discussed.


\setcounter{equation}{0}
\section{The Field Algebra}
\label{Algebra}


Here, we briefly recall the structure of the field algebra of
lattice QCD, for details we refer to \cite{KR}.

We consider QCD in the Hamiltonian framework on a finite regular
cubic lattice $\Lambda \subset \mathbb{Z}^3$, with  $\mathbb{Z}^3$
being the infinite regular lattice in 3 dimensions. We denote the
lattice boundary by $\partial \Lambda$ and the set of oriented,
$j$-dimensional elements of $\Lambda$, respectively $\partial
\Lambda$, by ${\Lambda}^j$, respectively $\partial \Lambda^j$,
where $j = 0,1,2,3 \, .$ Such elements are (in increasing order of
$j$) called sites, links, plaquettes and cubes. Moreover, we denote
the set of external links connecting boundary sites of $\Lambda$
with ``the rest of the world'' by $\Lambda^1_{\infty}$ and the set
of endpoints of external links at infinity by
$\Lambda^0_{\infty}$. For the purposes of this paper, we may
assume that for each boundary site there is exactly one link with
infinity. Then, external links are labelled by boundary sites and
we can denote them by $(x,\infty )$ with $x \in \partial
\Lambda^0$. The set of non-oriented $j$-dimensional elements will
be denoted by $|{\Lambda}|^j$. If, for instance, $(x,y) \in
{\Lambda}^1$  is an oriented link, then by $|(x,y)| \in
|{\Lambda}|^1$ we mean the corresponding non-oriented link. The
same notation applies to $\partial \Lambda$ and $\Lambda_{\infty}
\, .$

The basic fields of lattice QCD are quarks living at lattice sites
and gluons living on links, including links connecting the lattice
under consideration with ``infinity''. The field algebra is thus,
by definition, the tensor product of fermionic and bosonic
algebras:
\begin{equation}
\label{fieldalgebra}
{\mathfrak A}_{\Lambda} :=
{\mathfrak F}_{\Lambda} \otimes {\mathfrak B}_{\Lambda} \, ,
\end{equation}
with
\begin{equation}
\label{fermifieldalgebra}
{\mathfrak F}_{\Lambda} :=  \bigotimes_{x \in
|\Lambda |^0} {\mathfrak F}_x
\end{equation}
and
\begin{equation}
\label{bosonfieldalgebra}
{\mathfrak B}_{\Lambda} := {\mathfrak B}_{\Lambda}^i \otimes
{\mathfrak B}_{\Lambda}^b =
 \bigotimes_{|(x,y)| \in |\Lambda |^1} {\mathfrak B}_{|(x,y)|}
\bigotimes_{x \in \partial \Lambda} {\mathfrak B}_{|(x,\infty)|} \, .
\end{equation}
Here, ${\mathfrak B}_{\Lambda}^i$ and ${\mathfrak B}_{\Lambda}^b$
are the internal and boundary bosonic algebras respectively. We
impose {\em locality} of the lattice quantum fields by {\em
postulating} that the algebras corresponding to different elements
of $\Lambda$ commute with each other.

The fermionic field algebra ${\mathfrak F}_x$ associated with the
lattice site $x$ is the algebra of canonical anticommutation
relations (CAR) of quarks at $x$. The quark field generators are
denoted by
\begin{equation}
|{\Lambda}|^0 \ni x \rightarrow  {\psi}^{aA}(x) \in {\mathfrak
F}_x \, ,
\end{equation}
where $a$ stands for bispinorial and (possibly) flavour degrees of
freedom and $A = 1,2,3$ is the colour index corresponding to the
fundamental representation of the gauge group $G = SU(3)$. (In
what follows, writing $G$ we have in mind $SU(3)$, but essentially
our discussion can be extended to arbitrary compact groups and
their representations.) The conjugate quark field is denoted by
${\psi^{*}}_{aA}(x)  \, .$ The only nontrivial canonical
anti-commutation relations for generators of ${\mathfrak F}_x$
read:
\begin{equation}
\label{CCR3}
[{\psi^{*}}_{aA}(x), \psi^{bB}(x)]_+ = {\delta^B}_A
{\delta^b}_a \, .
\end{equation}

The bosonic field algebra ${\mathfrak B}_{|(x,y)|}$ associated
with the non-oriented link $|(x,y)|$, (where $y$ also stands for
$\infty$), is given in terms of its isomorphic copies ${\mathfrak
B}_{(x,y)}$ and ${\mathfrak B}_{(y,x)}$, corresponding to the two
orientations of the link $(x,y)\, .$ The algebra ${\mathfrak
B}_{(x,y)}$ is generated by matrix elements of the gluonic gauge
potential on the link $(x,y) \, ,$
\begin{equation}
{\Lambda}^1 \ni (x,y) \rightarrow  {U^A}_B(x,y) \in  {\mathfrak
C}_{(x,y)} \, ,
\end{equation}
with ${\mathfrak C}_{(x,y)} \cong C(G)$ being the commutative
$C^*$-algebra of continuous functions on $G$ and $A,B = 1,2,3$
denoting colour indices, and by colour electric fields, 
\begin{equation}
{\Lambda}^1 \ni (x,y) \rightarrow  {E^A}_B(x,y)  \in
{\mathfrak g}_{(x,y)} \, ,
\end{equation}
spanning the Lie algebra
${\mathfrak g}_{(x,y)} \cong  su(3)$. 
These elements generate, in the sense of Woronowicz \cite{unb},
the $C^*$-subalgebra ${\mathfrak P}_{(x,y)} \cong C^*(G) \subset
{\mathfrak B}_{(x,y)}$.

Observe that $G$ acts on $C(G)$ naturally by left translations,
\begin{equation}
\label{left} \alpha_g(u)(g') := u(g^{-1} g') \, \, , \, \,
u \in C(G) \, .
\end{equation}
Differentiating this relation, we get an action of $e \in su(3)$
on $u \in C^{\infty}(G)$ by the corresponding right invariant
vector field $e^R$. Thus, we have a natural commutator between
generators of ${\mathfrak P}_{(x,y)}$ and smooth elements of
${\mathfrak C}_{(x,y)}:$
\begin{equation}
\label{commut-e-f} i \ [e , u ] :=  e^R(u)  \ .
\end{equation}

To summarize, we have a $C^*$-dynamical system $\left({\mathfrak
C}_{(x,y)},G,\alpha\right),$ with automorphism $\alpha$ given by
the left action (\ref{left}). The field algebra ${\mathfrak
B}_{(x,y)}$ is, by definition, the corresponding crossed product
$C^*$-algebra,
\begin{equation}
\label{balgebra-xy} {\mathfrak B}_{(x,y)} :=
 {\mathfrak C}_{(x,y)} \otimes_{\alpha} G \, .
\end{equation}
We refer to \cite{Ped,Brat} for these notions.

The transformation law of elements of ${\mathfrak B}_{(x,y)}$
under the change of the link orientation is derived from the fact
that the (classical) $G$-valued parallel transporter $g(x,y)$ on
$(x,y)$ transforms to $g^{-1}(x,y)$ under the change of
orientation. This transformation lifts naturally to an isomorphism
\begin{equation}
\label{izo-I} {\cal I}_{(x,y)} : {\mathfrak B}_{(x,y)} \rightarrow
{\mathfrak B}_{(y,x)} \,
\end{equation}
of field algebras, defined by:
\begin{equation}\label{1izo}
{\cal I}_{(x,y)}(e,f):= (\breve{e},\breve{f}) \ ,
\end{equation}
where $\breve{f}(g):= f(g^{-1})$ and $\breve{e}$ is the left
invariant vector field on $G$, generated by $-e$. The bosonic
field algebra ${\mathfrak B}_{|(x,y)|}$ is obtained from
${\mathfrak B}_{(x,y)}$ and ${\mathfrak B}_{(y,x)}$ by identifying
them via ${\cal I}_{(x,y)}$.

Next, we give a full list of relations satisfied by generators of
${\mathfrak B}_{|(x,y)|}$. Being functions on $SU(3)\, ,$ the
generators of ${\mathfrak C}_{(x,y)}$ have to fulfil   the
following conditions:
\begin{eqnarray}
\label{unitary1} ({U^A}_B(x,y))^* {U_A}^C(x,y) & = & {\delta^C}_B
\ {\bf 1} \ ,
\\
\epsilon_{ABC}\  {U^A}_D(x,y) {U^B}_E(x,y) {U^C}_F(x,y) & = & \
\epsilon_{DEF} \ {\bf 1} \ . \label{unitary2}
\end{eqnarray}
The entries of the colour electric field obviously fulfil
\begin{eqnarray}
\label{self-adj} ({E^A}_B(x,y))^* & = & {E_B}^A(x,y) \, .
\end{eqnarray}
The transformation law (\ref{1izo}) reads
\begin{eqnarray}\label{izo-U}
  {U^A}_B(y,x) &=& {{\breve U}^A}_{\ \ B}(x,y) = ({U_B}^A(x,y))^* \ ,
  \\
  \label{ER-EL-1}
  {E^A}_B(y,x)  &=&  {{\breve E}^A}_{\ \ B}(x,y) =
   - {U^A}_D(y,x) {U^C}_B(x,y) {E^D}_C(x,y)
\end{eqnarray}
and the $su(3)$-commutation relations take the form
\begin{equation}
\label{CCR2} [{E^A}_B(x,y) ,{E^C}_D(u,z)]  =  \delta_{xu}
\delta_{yz} \left({\delta^C}_B {E^A}_D(x,y)  - {\delta^A}_D
{E^C}_B(x,y) \right)  \ .
\end{equation}
Finally, the generalized canonical commutation relations
(\ref{commut-e-f}) are given by:
\begin{eqnarray}
\label{CCR1} i \ [{E^A}_B(x,y),{U^C}_D(u,z)] & = & + \delta_{xu}
\delta_{yz} \left({\delta^C}_B {U^A}_D(x,y) -\frac{1}{3}
{\delta^A}_B {U^C}_D(x,y) \right) \nonumber \\ & & - \delta_{xz}
\delta_{yu} \left({\delta^A}_D {U^C}_B(y,x) -\frac{1}{3}
{\delta^A}_B {U^C}_D(y,x) \right)  \, .
\end{eqnarray}

\vspace{0.2cm} \noindent To summarize, the field algebra
${\mathfrak A}_{\Lambda}$, given by (\ref{fieldalgebra}) --
(\ref{bosonfieldalgebra}), is a $C^*$-algebra, generated by
elements
\begin{equation}
\label{gen-F}
  \left\{ \psi^{aA}(x) \, , \, {\psi^{*}}_{aA}(x) \, ,
  \, {U^A}_B(x,y) \, , \, {E^A}_B(x,y)
  \right\} \, ,
\end{equation}
fulfilling relations (\ref{unitary1}) -- (\ref{ER-EL-1}), together
with canonical (anti-) commutation relations (\ref{CCR3}),
(\ref{CCR2}) and (\ref{CCR1}).

Using standard arguments, one can prove that the field algebra
${\mathfrak A}_{\Lambda}$ has a unique (up to unitary equivalence)
irreducible representation. This representation is obtained as
follows, for details see \cite{KR1}:
\begin{enumerate}
\item Take the representation $\pi$ of the commutative
$C^*$-algebra $C(G)$ given by multiplication with elements of
$C(G) \, .$ \item Consider the left regular (unitary)
representation $\hat \pi$ of $G$ on $L^2(G,\mu) \, ,$
\begin{equation}
\label{leftreg} ({\hat \pi}(g)\xi)(g') := \xi(g^{-1} g') \, \, , \, \,
\xi \in L^2(G,\mu) \, .
\end{equation}
One easily calculates
\begin{equation}
\label{commpi-pi-hat}
{\hat \pi}(g) \circ \pi(u) \circ {\hat
\pi}(g^{-1}) =  \pi \left( \alpha_g(u) \right) \, ,
\end{equation}
showing that the pair $(\pi, {\hat \pi})$ defines a {\em
covariant} representation of $(C(G),G,\alpha)$ on $L^2(G,\mu) \,
.$ Differentiating (\ref{commpi-pi-hat}) yields the generalized
canonical commutation relations (\ref{commut-e-f}).
\end{enumerate}

By the Gelfand-Najmark theorem for commutative $C^*$-algebras, we
have a spectral measure $dE$ on $G \, ,$ such that $ \pi(u) =
\int u(g) \, dE(g) \, , $ for $u \in C(G) \, .$ Next, equation
(\ref{commpi-pi-hat}) implies
\begin{equation}
\label{commpi-pi-hat2}
L_g \circ dE(g') \circ L_{g^{-1}} =  dE(g g')  \, ,
\end{equation}
showing that the spectral measure $dE$ defines a transitive system
of imprimitivity for the representation ${\hat \pi}$ of $G$ based
on the group manifold $G\, .$ Then, the imprimitivity theorem, see
\cite{BaRa},\cite{Ki1}, yields uniqueness, up to unitary
equivalence. Finally, by the one-one-correspondence between
covariant representations of $C^*$-dynamical systems and
non-degenerate representations of the corresponding crossed
products we get a unique irreducible representation of $C(G)
\otimes_{\alpha} G$. Disregarding the $C^*$-context, this
statement is a classical result of Mackey, see \cite{Ki1} and
references therein. It generalizes the classical uniquess theorem
by von Neumann to the case of commutation relations
(\ref{commut-e-f}).

We take the tensor product of the above irreducible
representations over all links:
\begin{equation}
\label{productSchroedinger}
\bigotimes_{(x,y) \in \Lambda^1} L^2({\cal C}_{(x,y)},\mu)
\bigotimes_{x \in {\partial \Lambda}^0}
L^2({\cal C}_{(x,\infty)},\mu) \cong L^2({\cal C},\mu) \, , \,
\end{equation}
where
$$
{\cal C} :=  \prod_{(x,y) \in \Lambda^1} {\cal C}_{(x,y)} \,
\prod_{x \in {\partial \Lambda}^0} {\cal C}_{(x,\infty)} \, ,
$$
and each space ${\cal C}_{(x,y)}$ being diffeomorphic to the group
space $G\, .$ This is the unique representation space of the
gluonic field algebra ${\mathfrak B}_{\Lambda}$. Moreover, using
the classical uniqueness theorem for CAR-representations by Jordan
and Wigner, any representation of fermionic fields is equivalent
to the fermionic Fock representation. Finally, we have the
following isomorphism of $C^*$-algebras:
$$
C(G) \otimes_{\alpha} G \cong
\mathfrak{K}(L^2(G,\mu)) \, ,
$$
where $\mathfrak{K}(L^2(G,\mu))$ denotes the algebra of compact
operators on $L^2(G,\mu)$, for the proof see \cite{KR1}. This
implies that the field algebra ${\mathfrak A}_{\Lambda}$ can be
identified with the algebra ${\mathfrak K}(H_\Lambda)$ of compact
operators on the Hilbert space
\begin{equation}
 \label{irreps}
  H_\Lambda =
  {\cal F}({\mathbb C}^{12N}) \otimes
  L^2({\cal C},\mu)
  \, ,
\end{equation}
with ${\cal F}({\mathbb C}^{12N})$ denoting the fermionic Fock
space generated by $12N$ anti-commuting pairs of quark fields.


\setcounter{equation}{0}
\section{Gauge Transformations and Gauss Law}
\label{Gauge--Gauss}


The group $G_{\Lambda}$ of local gauge transformations related to
the lattice ${\Lambda}$ consists of mappings
$$
\Lambda^0 \ni x \rightarrow g(x) \in G \, ,
$$
which represent internal gauge transformations, and of gauge
transformations at infinity,
$$
\Lambda^0_{\infty} \ni z \rightarrow g(z) \in G \, .
$$
Thus,
\begin{equation}
\label{gaugegroup}
G_{\Lambda} := G_i \times G_{\infty}
= \prod_{x \in \Lambda^0} G_x \, \prod_{z \in \Lambda^0_{\infty}} G_z
\, ,
\end{equation}
with $ G_y \cong SU(3)\, , $ for every $y$.

The group $G_{\Lambda}$ acts on the classical configuration space
${\cal C}$ as follows:
$$
{\cal C}_{(x,y)} \ni g(x,y) \rightarrow g(x) g(x,y) g(y)^{-1} \in {\cal C}_{(x,y)} \, ,
$$
with  $g(x) \in G_x$ and $g(y) \in G_y$. This action lifts
naturally to functions on ${\cal C}$. Moreover, we have an action
of $G_x$ on itself by inner automorphisms. This yields an action
of $G_{\Lambda}$ by automorphisms on each $C^*$-dynamical system
$({\mathfrak C}_{(x,y)},G,\alpha)$ and, therefore, on the gluonic
field algebra ${\mathfrak B}_{\Lambda}$. For generators of
${\mathfrak B}_{(x,y)} \subset {\mathfrak B}_{\Lambda}$, this
action is given by
\begin{eqnarray}
\label{gluonic-gauge}
{U^A}_B(x,y) & \rightarrow & {g^A}_C(x)
{U^C}_D(x,y) {{(g^{-1})}^D}_B(y) \ ,
\\
{E^A}_B(x,y) & \rightarrow & {g^A}_C(x) {E^C}_D(x,y)
{{(g^{-1})}^D}_B(x) \ ,
\end{eqnarray}
with $y$ standing also for $\infty$.
Fermionic generators transform under the fundamental
representation:
\begin{equation}
\label{fermionic-gauge} \psi^{aA}(x)  \rightarrow  {g^A}_B(x)
\psi^{aB}(x) \ .
\end{equation}
To summarize, the group of local gauge transformation
$G_{\Lambda}$ acts on the field algebra ${\mathfrak A}_{\Lambda}$
in a natural way by automorphisms.

The local Gauss law at $x \in {\Lambda}^0 $
reads
\begin{equation}
 \label{Gauss}
\sum_{y \leftrightarrow x}
 {E^A}_B(x,y)  = {\rho^A}_B(x)\ ,
\end{equation}
where
\begin{equation}
\label{rho} {\rho^A}_B(x) = \sum_a \left( \psi^{*aA}(x)
{\psi^a}_{B}(x) - \frac 13 \delta ^A{}_B \psi^{*aC}(x)
{\psi^a}_{C}(x) \right) \
\end{equation}
is the local matter charge density, fulfilling ${\rho^A}_A(x) =
0$.

In \cite{KR} we have analyzed the Gauss law in detail. Here, we
briefly recall the main result. Consider any integrable
representation $F$ of the Lie algebra $su(3)$ on a Hilbert space
$\cal H \, ,$ i.~e.~a collection of operators ${F^A}_B$ in $\cal
H$, fulfilling ${F^A}_A = 0 \, ,$ $({F^A}_B)^*  =  {F_B}^A$ and
\begin{equation}
\label{CCRF}
[{F^A}_B , {F^C}_D] = {\delta^C}_B {F^A}_D  - {\delta^A}_D {F^C}_B  \, .
\end{equation}
By (\ref{CCR2}), (\ref{rho}) and (\ref{CCR3}), the operators
${E^A}_B(x,y)$ and ${\rho^A}_B(x)$, occurring on both sides of the
local Gauss law, are of this type. Integrability means that for
each $F$ there exists a unitary representation $SU(3) \ni g
\rightarrow {\bar F}(g) \in B({\cal H})$ of the group $SU(3)\, .$
If $F$ and $G$ are two commuting (integrable) representations of
$su(3)$, then so is $F+G$. Moreover, $-F^*$ is also a
representation of $su(3) \, .$ Such a collection of operators is
an {\em operator domain} in the sense of Woronowicz (see
\cite{kot}).

We define an operator function on this domain, i.~e.~a mapping $F
\rightarrow \varphi(F)\, ,$ which satisfies $\varphi(U F U^{-1} )
= U \varphi(F) U^{-1}$ for an arbitrary isometry $U$, as follows:
For any integrable representation $F$ of $su(3)$, consider the
corresponding representation ${\bar F}$ of $SU(3)$. Its
restriction to the center ${\cal Z}$ of $SU(3)$ acts as a multiple
of the identity on each irreducible subspace ${\cal H}_\alpha$ of
${\bar F} \, ,$
$$
{\bar F}(c) |_{{\cal H}_\alpha} = \chi_{\bar F}^{\alpha}(c)
\cdot {\bf 1}_{{\cal H}_\alpha} \, \, ,
\, \, \, c \in {\cal Z} \, .
$$
Obviously, $\chi_{\bar F}^{\alpha}$ is a character on ${\cal Z}$
and, therefore, $(\chi_{\bar F}^{\alpha}(c))^3 = 1$. Since
the group of characters on ${\cal Z} =\{ \zeta \cdot {\bf 1}_3 \ |
\  \zeta^3 = 1 \ , \ \zeta \in {\mathbb C} \}$ is isomorphic to the additive
group ${\mathbb Z}_3 \cong \{-1,0,1\}$, there exists a  ${\mathbb Z}_3$-valued operator function
$F \rightarrow \varphi(F)$, defined by
\begin{equation}
\label{innephi}
\zeta^{\, \varphi_{\alpha} (F)} = \chi_{\bar F}^{\alpha}
(\, \zeta \cdot {\bf 1}_3) \, , \quad
\varphi(F) = \sum_{\alpha} \varphi_{\alpha} (F) \,
{\bf 1}_{{\cal H}_\alpha} \, .
\end{equation}
Since $\chi_{\bar F}^{\alpha} $  are characters, we have
\begin{equation}
\label{fi3}
\varphi(F + G)  =   \varphi(F) + \varphi(G) \, ,
\end{equation}
for $F$ and $G$ commuting. Now, using the equivalence of the
irreducible representation ${\bar F}$ of $SU(3)$ with highest
weight $(m,n)$ with the tensor representation in the space
${{\mathbb T}}^m{}_n({\mathbb C}^3)$ of $m$-contravariant,
$n$-covariant, completely symmetric and traceless tensors over
${\mathbb C}^3$, we get
\begin{equation}
\label{innephi2}
\chi_{\bar F}^{\alpha}(z) = \zeta^{\, \varphi_{\alpha} (F)} =
\zeta^{\, m(\alpha) - n(\alpha)} \, ,
\end{equation}
for $z = \zeta \cdot {\bf 1}_3 \in {\cal Z} \, .$ Thus, we have
\begin{equation}
\label{varphi--m-n}
\varphi_{\alpha} (F) = (m(\alpha)-n(\alpha)) \,\,\, \text{mod} \,\,\, 3
\end{equation}
for every irreducible highest weight representation
$(m(\alpha),n(\alpha)) \, .$ In \cite{KR} we have given an
explicit construction of $\varphi (F)$ in terms of Casimir
operators of $F \, .$

Applying $\varphi$ to the local Gauss law (\ref{Gauss}) and using
additivity (\ref{fi3}) we obtain a gauge invariant equation for
operators with eigenvalues in ${\mathbb Z}_3$:
\begin{equation}
\label{inGauss}
\sum_y \varphi(E(x,y)) = \varphi(\rho(x)) \ ,
\end{equation}
valid at every lattice site $x$. The quantity on the right hand
side is the (gauge invariant) local colour charge density carried
by the quark field. Using the transformation law (\ref{ER-EL-1})
for $E(x,y)$ under the change of the link orientation and
additivity (\ref{fi3}) of $\varphi$, one can show that
\begin{equation}
\varphi(E(x,y)) + \varphi(E(y,x)) = 0  \, ,
\end{equation}
for every lattice bond $(x,y)$. Now we take the sum of equations
(\ref{inGauss}) over all lattice sites $x \in \Lambda$. Due to the
above identity, all terms on the left hand side cancel, except for
contributions coming from the boundary. This way we obtain the
total flux through the boundary $\partial \Lambda$ of $\Lambda$:
\begin{equation}\label{flux}
  \Phi_{\partial \Lambda} := \sum_{x \in \partial \Lambda}
  \varphi(E(x,\infty )) \ ,
\end{equation}
On the right hand side we get the (gauge invariant) global colour
charge ({\em triality}), carried by the matter field
\begin{equation}
\label{globalM}
{\mathfrak{t}}_{\Lambda} := \sum_{x \in \Lambda}
\varphi(\rho(x)) \, .
\end{equation}
Both quantities appearing in the global Gauss law
\begin{equation}\label{gG}
  \Phi_{\partial \Lambda} = {\mathfrak{t}}_{\Lambda} \ ,
\end{equation}
take values in the center ${\mathbb Z}_3$ of $SU(3)$.


\setcounter{equation}{0}
\section{The Observable Algebra}
\label{Observablealgebra}



\subsection{The Algebra of Internal Observables }
\label{int--Observablealgebra}


\noindent Physical observables, internal relative to $\Lambda$
are, by definition, gauge invariant fields, respecting the Gauss
law. Hence, we proceed as follows:
\begin{enumerate}
\item
Take the subalgebra
$
  {\mathfrak A}^{G_{\Lambda}}
  \subset {\mathfrak A}_{\Lambda}
$
of $G_{\Lambda}$-invariant elements of ${\mathfrak A}_{\Lambda}$.
\item
Require vanishing of the ideal ${\mathfrak I}^i_{\Lambda} \cap
    {\mathfrak A}^{G_{\Lambda}} \, ,$
generated by local Gauss laws at all lattice sites.
\end{enumerate}
Then, the algebra of internal observables, relative to $\Lambda\,
,$ is given by:
$$
{\mathfrak O}^i_{\Lambda} = {\mathfrak A}^{G_{\Lambda}} /
    \{{\mathfrak I}^i_{\Lambda} \cap
    {\mathfrak A}^{G_{\Lambda}} \} \, .
$$
Using the identification ${\mathfrak A}_{\Lambda} \cong {\mathfrak
K}(H_{\Lambda})$ yields a unitary representation of $G_{\Lambda}$
in ${\mathfrak K}(H_{\Lambda})\, .$ Thus, the subalgebra
${\mathfrak A}^{G_{\Lambda}}$ can be viewed as the commutant
$(G_{\Lambda})'$ of this representation in ${\mathfrak
K}(H_{\Lambda})$. Consider the $G_\Lambda^i$-invariant subspace
$$
  {\cal H}_{\Lambda} := \{ h \in H_{\Lambda} \, \, | \, \,
  G_\Lambda^i h = h \, \} \ .
$$
\begin{Theorem}
The algebra of internal observables is canonically isomorphic with
the algebra of those compact operators on the Hilbert space ${\cal
H}_{\Lambda}$, which commute with the action of the group
$G_\Lambda^{\infty}$,
$$
  {\mathfrak O}^i_{\Lambda} \cong {\mathfrak K}({\cal
  H}_{\Lambda}) \cap    (G_\Lambda^{\infty})' \ .
$$
\end{Theorem}
For the proof see \cite{KR1}.
\noindent
Next, we want to classify the irreducible representations of
${\mathfrak O}^i_{\Lambda}$:\\
Note that the restriction of the action of $G^{\infty}_{\Lambda}$
to ${\cal H}_{\Lambda}$ is not irreducible. Thus,  ${\cal
H}_{\Lambda}$ splits into the direct sum of irreducible subspaces
of $G^{\infty}_{\Lambda} \, .$ These are labelled by sequences of
highest weights,
$$
  ({\bf m}, {\bf n}) = (m_{z_1}, \dots , m_{z_M}
  ; n_{z_1}, \dots , n_{z_M}) \ , \, z_i \in
\Lambda^0_{\infty} \, ,
$$
describing the boundary flux distributions, carried by the gluonic
field. We decompose
$$
{\cal H}_{\Lambda} =
\bigoplus {\cal H}^{({\bf m}, {\bf n})}_{\Lambda} \ ,
$$
with ${\cal H}^{({\bf m}, {\bf n})}_{\Lambda}$ denoting the sum of
all irreducible subspaces with respect to the action of
$G^{\infty}_{\Lambda}$, carrying the same type $({\bf m}, {\bf
n})$. Then we have
$$
  {\mathfrak O}^i_{\Lambda}
  {\cal H}^{({\bf m}, {\bf n})}_{\Lambda}
  \subset
  {\cal H}^{({\bf m}, {\bf n})}_{\Lambda} \ .
$$

\begin{Theorem}
The irreducible representations of ${\mathfrak O}^i_{\Lambda}$ are
labelled by highest weight representations $({\bf m}, {\bf n})$ of
$G_\Lambda^{\infty}$.  For any $({\bf m}, {\bf n})$, the
corresponding irreducible representation of ${\mathfrak
O}^i_{\Lambda}$ coincides with the algebra of those compact
operators on ${\cal H}_{\Lambda}^{({\bf m}, {\bf n})}$, which
commute with the action of the group $G_\Lambda^{\infty}$.
\end{Theorem}
\noindent
For the proof see \cite{KR1}.


\subsection{The Full Algebra of Observables }
\label{full--Observablealgebra}


For constructing the thermodynamical limit of finite lattice QCD,
one has to take into account ``correlations with the rest of the
world'': Consider two lattices $\Lambda_1$ and $\Lambda_2$, having
a common wall and denote $\widetilde\Lambda = \Lambda_1 \cup
\Lambda_2 \, .$ If $u \in \Lambda_1$ and $v \in \Lambda_2$ are
adjacent points in $\widetilde \Lambda$, we identify their
infinities. Imagine this joint infinity $z$ as the middle point of
the connecting link $(u,v) \, .$ We put:
$$
U^A{}_{B}(u,v) := U^A{}_{C}(u,z) U^C{}_{B}(z,v) .
$$
Now, typical observables describing correlations between
$\Lambda_1$ and $\Lambda_2$ are:
$$
J^{a b}_{\gamma}(x,y) := {\psi^{*a}}_{A}(x) \, U^A_{\gamma \, B} \,
\psi^{bB}(y) \, , \quad  U_{\tau} := U^A_{\tau \, A} \, ,
$$
with $\gamma$ being a path starting at some point $x \in
\Lambda_1$ and ending at the point $y \in \Lambda_2 \, ,$ and
$\tau$ being a closed path running partially through $\Lambda_1$
and partially through $\Lambda_2 \, .$ (Strictly speaking, these
are elements of the multiplier algebra of the observable algebra
${\mathfrak O}^i_{\widetilde\Lambda}\, .$)

In order to construct observables of this type, we have to
admit ``charge carrying'' fields, ``having free
tensor indices at infinity'', like ${\psi^{*a}}_{A}(x) \,
U^A_{\gamma_1 \, C}(x,z)$ and $U^C_{\gamma_2 \, B}(z,y) \,
\psi^{bB}(y)$, with $z$ being a joint infinity point.

Thus, we are naturally led to extend the Hilbert space ${\cal
H}_{\Lambda}$ to ${\mathbb T}_{\infty} \otimes {\cal H}_{\Lambda}
$ by tensorising it with
$$
   {\mathbb T}_{\infty} :=
   \bigotimes_{z \in \Lambda^0_\infty} {\mathbb T}(z) \, \, \,  ,
\, \, \,  {\mathbb T}(z) :=
    \bigoplus_{(m,n)} {\mathbb T}^{m}_{n}(z) \  .
$$
Moreover, we extend both the action of $G_\Lambda^{\infty}$ on
${\cal H}_{\Lambda}$,
$$
    T(g) \left( t \otimes \Psi \right) :=
    t \otimes  \left( g \cdot \Psi \right)\ \,   ,
$$
and the action of $G_\Lambda^{\infty}$ on ${\mathbb T}_{\infty}$,
$$
    R(g) \left( t \otimes \Psi \right) :=
    \left( g \cdot t\right)  \otimes
    \Psi  \ .
$$
Tensors $t(z)$ occurring in $t \otimes \Psi  \in {\mathbb
T}_{\infty} \otimes {\cal H}_{\Lambda}$ represent ``quantum
averages'' over external degrees of freedom. Finally, we implement
one additional requirement: We demand compatibility of averaging
with gauging. This yields the following physical Hilbert space:
$$
{\bf H}_{\Lambda} :=
   \left\{ {\boldsymbol \Psi} \in {\mathbb T}_{\infty}
   \otimes {\cal H}_{\Lambda} \ | \  T(g)
   {\boldsymbol \Psi} = R(g)  {\boldsymbol \Psi}\, , \ \
    g \in G_\Lambda^{\infty}  \right\} \ .
$$
Consequently, the full algebra of observables is:
\begin{equation}
\label{fullobservalg}
  {\mathfrak O}_{\Lambda} :=
  {\mathfrak K}({\bf H}_{\Lambda}) \cap
  (G^{\infty}_{\Lambda})^\prime \ .
\end{equation}

Decompose
$$
  {\bf H}_{\Lambda} =
  \bigoplus_{({\bf m}, {\bf n}) }
  {\bf H}_{\Lambda}^{({\bf m}, {\bf n})}  \ ,
$$
with ${\bf H}_{\Lambda}^{({\bf m}, {\bf n})}$ denoting the
intersection of ${\bf H}_{\Lambda}$ with ${\mathbb
T}_{\infty}^{({\bf m}, {\bf n})} \otimes {\cal H}_{\Lambda}$
and
$$
   {\mathbb T}_{\infty}^{({\bf m}, {\bf n})} :=
   \bigotimes_{z \in \Lambda^0_\infty} {\mathbb T}^{m_z}_{n_z}(z) \, .
$$
We denote by ${\mathfrak O}^{\infty}_{\Lambda}$ the algebra of
compact operators acting on ${\mathbb T}_{\infty}$, invariant with
respect to the action of $G^{\infty}_{\Lambda}$. By classical
invariant theory, this algebra is generated (in the sense of
Woronowicz) by operations of tensorizing or contracting with
$SU(3)$-invariant tensors $\delta^A{}_B$, $\epsilon^{ABC}$ and
$\epsilon_{ABC}$, and by projection operators $P^{({\bf m}, {\bf
n})}$ onto ${\mathbb T}_{\infty}^{({\bf m}, {\bf n})} \subset
{\mathbb T}_{\infty}$. One shows:
$$
{\mathfrak O}_{\Lambda}  \cong {\mathfrak O}^i_{\Lambda}
\otimes {\mathfrak O}^{\infty}_{\Lambda} \, .
$$
The action of ${\mathfrak O}^{\infty}_{\Lambda}$ on ${\mathbb
T}_{\infty}$ is not irreducible, the image of $t(z) \in {\mathbb
T}^m_n(z)$ under this action is a sum of components belonging to
${\mathbb T}^k_l(z) \, ,$ with $(k,l)$ fulfilling
$$
(m - n)\ \mbox{\rm mod}\ 3 = (k - l)\
\mbox{\rm mod}\ 3\, .
$$
Thus, the ${\mathbb Z}_3$-valued flux
$$
  {\Phi}(z):=(m_z - n_z)\ \mbox{\rm mod}\ 3
$$
through each external link $(x,z)$, $z \in \Lambda^0_{\infty}$, is
conserved under the action of ${\mathfrak O}^{\infty}_{\Lambda}$.
We denote the sequence of ${\mathbb Z}_3$-valued fluxes assigned
to all boundary points by:
$$
\boldsymbol\Phi := (\Phi(z_1) , \Phi(z_2), \dots )
$$
and put
$$
  {\bf H}_{\Lambda}^{\boldsymbol\Phi} :=
  \bigoplus_{m(z_i) - n(z_i)\ {\rm mod} \ 3 \, = \, \Phi(z_i) }
{\bf H}_{\Lambda}^{({\bf m}, {\bf n})} \, .
$$
Then, obviously
$$
  {\bf H}_{\Lambda} = \bigoplus_{\boldsymbol\Phi}
  {\bf H}_{\Lambda}^{\boldsymbol\Phi} \ .
$$
\begin{Lemma}
\label{sectors-loc-triality} The spaces ${\bf
H}_{\Lambda}^{\boldsymbol\Phi}$ provide all the irreducible
representations of the algebra of observables ${\mathfrak
O}_{\Lambda}$.
\end{Lemma}
\noindent
The global flux associated with a given boundary flux
distribution $\boldsymbol\Phi$ :
$$
  \Phi_{\partial\Lambda} :=  \sum_{z \in \Lambda^0_\infty}
  \Phi(z) \quad \mbox{\rm mod}\ 3  \ .
$$
If we denote the total number of gluonic and antigluonic flux
lines running through the boundary by
$$
  m := \sum_{z_i \in \Lambda^0_\infty} m(z_i) \ , \
  n := \sum_{z_i \in \Lambda^0_\infty} n(z_i) \ ,
$$
then we have:
$$
\Phi_{\partial\Lambda} = (m-n) \, \, \mbox{\rm mod}\ 3 \, .
$$
\begin{Lemma}
\label{sectors-triality} The irreducible representations of
${\mathfrak O}_{\Lambda}$ in ${\bf H}_{\Lambda}^{\boldsymbol\Phi}$
and in ${\bf H}_{\Lambda}^{\boldsymbol\Phi^\prime}$ are unitarily
equivalent, if and only if ${\boldsymbol\Phi}$ and
${\boldsymbol\Phi^\prime}$ carry the same global flux,
$$
\Phi_{\partial\Lambda}=\Phi^\prime_{\partial\Lambda} \, .
$$
\end{Lemma}
This yields the following classification of irreducible
representations:
\begin{Theorem}
\label{glob-flux}
There are three inequivalent representations of
${\mathfrak O}_{\Lambda}$ labelled by values of the global flux
$\Phi_{\partial\Lambda}$. Consequently, the space ${\bf
H}_{\Lambda}$ splits into the sum of three eigenspaces of
$\Phi_{\partial\Lambda}$
\[
   {\bf H}_{\Lambda} = \bigoplus_{\lambda= -1,0,1}
   {\bf H}_{\Lambda}^\lambda \ .
\]
Each of the spaces ${\bf H}_{\Lambda}^\lambda$ is a sum of
superselection sectors ${\bf H}_{\Lambda}^{\boldsymbol\Phi}$
corresponding to all possible distributions ${\boldsymbol\Phi}$ of
the global flux $\lambda$. They carry equivalent representations
of ${\mathfrak O}_{\Lambda}$.
\end{Theorem}
\noindent
Finally, by the global Gauss law, we get
\begin{Corollary}
\label{glob-triality} The inequivalent representations of
${\mathfrak O}_{\Lambda}$ are labelled by eigenvalues of global
colour charge ${\mathfrak{t}}_{\Lambda}$.
\end{Corollary}

For the proof of the above statements we refer to \cite{KR1}.


\subsection{Generators and relations}
\label{generators-relations}


We recall the decomposition \eqref{fullobservalg},
\begin{equation}
\label{O=OxO1}
{\mathfrak O}_{\Lambda}  \cong {\mathfrak O}^i_{\Lambda}
\otimes {\mathfrak O}^{\infty}_{\Lambda} \, .
\end{equation}
As already mentioned above, ${\mathfrak O}^{\infty}_{\Lambda}$ is
generated by operations of tensorizing or contracting with
$SU(3)$-invariant tensors $\delta^A{}_B$, $\epsilon^{ABC}$ and
$\epsilon_{ABC}$, and by projection operators $P^{({\bf m}, {\bf
n})}$ onto ${\mathbb T}_{\infty}^{({\bf m}, {\bf n})} \subset
{\mathbb T}_{\infty}$. Thus, one has to find a complete set of
generators of ${\mathfrak O}^i_{\Lambda}\, .$ In \cite{KR1} we
have shown the following
\begin{Theorem}
\label{generators} The observable algebra ${\mathfrak
O}^i_{\Lambda}$ is generated by the following gauge invariant
elements (together with their conjugates):
\begin{eqnarray}
   \label{observableU}
   U_{\gamma} & := & U^A_{\gamma \, A} \, \\
   \label{observableE} E_{\gamma}(x,y) & := & U^A_{\gamma \, B} \,
   {E^B}_A(x,y) \, \\
   \label{observableJ} J^{a b}_{\gamma}(x,y) & :=
   & {\psi^{*a}}_{A}(x) \, U^A_{\gamma \, B} \, \psi^{bB}(y) \, \\
   \label{observableW} W^{a b c}_{\alpha \beta \gamma}(x,y,z) &:= &
   \tfrac{1}{6} \epsilon_{ABC} \, U^A_{\alpha \, D} \, U^B_{\beta \,
   E} \, U^C_{\gamma \, F} \, \psi^{aD}(x) \,\psi^{bE}(y)
   \,\psi^{cF}(z) \, ,
\end{eqnarray}
with $\gamma$ denoting an arbitrary closed lattice path in formula
(\ref{observableU}), a closed lattice path starting and ending at
$x$ in (\ref{observableE}) and a path from $x$ to $y$ in
(\ref{observableJ}). In formula (\ref{observableW}), $\alpha$,
$\beta$ and $\gamma$ are paths starting at some reference point
$t$ and ending at $x$, $y$ and $z$, respectively. In formula
(\ref{observableE}), both $x$ and $y$ stand also for $\infty$.
\end{Theorem}
Note that the observables $J^{a b}_{\gamma}$ and $W^{a b
c}_{\alpha \beta \gamma}$ represent hadronic matter of mesonic and
baryonic type.

The above set of generators is, however, highly redundant. In a
first step, it can be reduced by using the concept of a lattice
tree. As a result, one obtains a presentation of the observable
algebra in terms of tree data, which are still subject to gauge
transformations at the (arbitrarily chosen) tree root. Finally,
this gauge freedom has to be removed. This reduction procedure has
been discussed in detail in \cite{KR1}. The second step leads to
delicate problems (Gribov problem and the occurence of nongeneric
strata), which suggest that one should investigate the stratified
structure of the underlying classical configuration (resp. phase)
space in more detail. See \cite{ChKRS} for first results.

As a result of this reduction, the observable algebra ${\mathfrak
O}_{\Lambda}$ is obtained as
\begin{equation}
\label{tensor}
   {\mathfrak O}_{\Lambda}  = {\mathfrak O}^{glu}_{\tree} \otimes
   {\mathfrak O}^{mat}_{\tree} \otimes {\mathfrak O}^b_\Lambda
   \otimes {\mathfrak O}^\infty_\Lambda  \ .
\end{equation}
Here, the {\em gluonic} component ${\mathfrak O}^{glu}_{\tree}$
is generated by reduced gluonic tree data
$({\mathfrak{u}}_i,{\mathfrak{e}}_i)\,,$ $ i = 0, \dots , K-2 \,
,$ with $K$ denoting the number of off-tree lattice links. These
bosonic generators satisfy the generalized canonical commutation
relations over $G :$
\begin{eqnarray}
\label{e-e}
  \left[ {\mathfrak{e}}_i^r{}_s  ,{\mathfrak{e}}_j^p{}_q
  \right]
  &=&  \delta_{ij}\left({\delta^p}_s {\mathfrak{e}}_i^r{}_q  -
  {\delta^r}_q {\mathfrak{e}}_i^p{}_s \right)  \ , \\
  \label{e-u}
  \left[{\mathfrak{e}}_i^r{}_s ,{\mathfrak{u}}_j^p{}_q
  \right]
  & = &  \delta_{ij}
    \left({\delta^p}_s {\mathfrak{u}}_i^r{}_q - \tfrac{1}{3}
    {\delta^r}_s {\mathfrak{u}}_i^p{}_q \right) \, ,
    \\
    \left[{\mathfrak{u}}_i^r{}_s ,{\mathfrak{u}}_j^p{}_q
  \right]
  & = & 0
    \ ,\label{u-u}
\end{eqnarray}
with $r,s, \dots = 1,2,3 \, .$ The generators
$({\mathfrak{u}}_i,{\mathfrak{e}}_i)$ are subject to a certain
discrete symmetry described in \cite{KR1}, (which, however, is not
a remainder of the gauge symmmetry).

Applying the above gauge fixing procedure to the fermionic matter
field, we obtain fermionic operators $\phi_k$. Here, $k= (a,r,x)$
is a multi-index running from $1$ to $12 N \, ,$ with $N$ being
the number of lattice sites. These quantities fulfil the canonical
anti-commutation relations
\begin{equation}
\label{anti-comm--a}
\left[\phi^{*k} , \phi_l
\right]_+ = \delta^k{}_l \ .
\end{equation}
(We stress that in \cite{KR1}, the generators  $\phi_k$ were
denoted by ${\mathfrak{a}}_k \,.$ ) Again, an additional discrete
symmetry arises, because the gauge-fixing is defined only up to
the stabilizer  ${\mathbb Z}_3$ of the generic stratum of the
underlying classical configuration space. Thus, strictly speaking,
the generators $\phi_k$ {\em are not observables}, whereas the
bosonic quantities ${\mathfrak{u}}$ and ${\mathfrak{e}}$ are,
because they are not affected by this ambiguity. It is clear from
classical invariant theory that only the following combinations of
$\phi^{*k}$ and $\phi_k$ (together with functions built from them)
are observables:
\begin{eqnarray}
\mathfrak{j}^k{}_l & = & \phi^{*k} \, \phi_l \, ,
\label{observ-a1}\\
\mathfrak{i}^k{}_l& = & \phi_l \, \phi^{*k}
\label{observ-a12} \, ,\\
\mathfrak{w}_{pqr} & = & \phi_p \, \phi_q \,
\phi_r \, ,
\label{observ-a2} \\
{\mathfrak{w}}^*{}^{ijk} & = & \phi^{*k} \,
\phi^{*j} \, \phi^{*i} \,.
\label{observ-a3}
\end{eqnarray}
We have, of course,
\begin{equation}
\label{1-j1}
\mathfrak{i}^k{}_l = \delta^k{}_l {\bf 1} - \mathfrak{j}^k{}_l\, .
\end{equation}
Thus, the matter field component ${\mathfrak O}^{mat}_{\tree}$ is
generated by the set $\{\mathfrak{j}, \mathfrak{w},
{\mathfrak{w}}^*\} \, ,$ together with the unit element ${\bf 1}\,
.$ These generators are observables of hadronic type. Thus, in
what follows we call ${\mathfrak O}^{mat}_{\tree}$ {\em hadronic}
component of the observable algebra, or simply {\em hadronic
subalgebra}.

Finally, ${\mathfrak O}^b_\Lambda$ is generated by (gauge invariant)
color electric boundary fluxes and the generators of
${\mathfrak O}^\infty_\Lambda $ have been already given above.

By the uniqueness theorem for generalized CCR, fulfilled by
generators $({\mathfrak{u}}_i,{\mathfrak{e}}_i)\,,$ of the gluonic
subalgebra ${\mathfrak O}^{glu}_{\tree}\, ,$ the problem of
classifying irreducible representations of ${\mathfrak
O}_{\Lambda}$ is reduced to classifying irreducible
representations of the hadronic subalgebra ${\mathfrak
O}^{mat}_{\tree} \, .$ For that purpose, the structure of this
algebra will be investigated in the sequel.

\setcounter{equation}{0}
\section{Structure of the Hadronic Subalgebra}



\subsection{Generators and Relations}
\label{gen--rel}


We start analyzing ${\mathfrak O}^{mat}_{\tree}$ by listing
relations implied from definitions \eqref{observ-a1} --
\eqref{observ-a3}.

First, note that ${\mathfrak O}^{mat}_{\tree}$ is a unital
$*$-algebra, with unit element ${\bf 1}$ and $*$-operation given by
\begin{eqnarray}
  \left(\mathfrak{j}^k{}_l \right)^* & = & \mathfrak{j}^l{}_k \ ,
  \label{*-struct1}\\
  \label{*-struct}
  \left(\mathfrak{w}_{pqr}\right)^* & = &  {\mathfrak{w}}^{*rqp}
  \, , \\
  \left({\mathfrak{w}}^{*ijk}\right)^* & =& \mathfrak{w}_{ijk}
  \, , \label{*-struct3}
\end{eqnarray}
where the generators $\mathfrak{w}$ and ${\mathfrak{w}}^*$  are
totally antisymmetric in their indices:
\begin{equation}
     \mathfrak{w}_{m k n}=\mathfrak{w}_{k n m} =
  \mathfrak{w}_{n m k} = - \mathfrak{w}_{k m n} \ .
  \label{skewsymm1}
\end{equation}

Next, the anticommutation relations (\ref{anti-comm--a})
immediately yield:
\begin{equation}
\label{commrel-j}
  \left[ \mathfrak{j}^k{}_l , \mathfrak{j}^m{}_n \right]  =
  \delta^m{}_l \, \mathfrak{j}^k{}_n -
  \delta^k{}_n \, \mathfrak{j}^m{}_l \, ,
\end{equation}
\begin{equation}
\label{commrel-jw} \left[ \mathfrak{j}^i{}_k , \mathfrak{w}_{l m
 n} \right]  =
 - \delta^i{}_l \, \mathfrak{w}_{k m n }
 - \delta^i{}_m \, \mathfrak{w}_{l k n}
 - \delta^i{}_n \, \mathfrak{w}_{l m k} \, ,
\end{equation}
and, consequently,
\begin{equation}
\label{commrel-jw*}
    \left[ \mathfrak{j}^k{}_i , \mathfrak{w}^{*l m n} \right] =
   \delta^l{}_i \, \mathfrak{w}^{*k m n }
   + \delta^m{}_i \, \mathfrak{w}^{*l k n}
   + \delta^n{}_i \, \mathfrak{w}^{*l m k} \ .
\end{equation}
Observe that \eqref{commrel-j} are the commutation relations of
$gl( n,\mathbb{C})$, with $n = 12N$. As a direct consequence of
\eqref{observ-a1}, these generators fulfil a number of additional
quadratic relations:\\ First, the diagonal generators
$\mathfrak{j}^k{}_k$ are idempotent
\begin{equation}
\label{idempotent1}
\left(\mathfrak{j}^k{}_k \right)^2 = \mathfrak{j}^k{}_k \ .
\end{equation}
Because of the Hermicity condition \eqref{*-struct1} they are,
thus, projectors. Commutation relations \eqref{commrel-j} implies
that they all commute with each other.

Finally, products of $\mathfrak{w}$ and $\mathfrak{w}^*$ can be
expressed in terms of $\mathfrak{j}$'s:
\begin{eqnarray}
  \mathfrak{w}^{*k m n } \mathfrak{w}_{k m n} &=&
  \mathfrak{j}^k{}_k \mathfrak{j}^m{}_m \mathfrak{j}^n{}_n \, ,
  \ \ \ \ \mbox{\rm for different } k,m,n \ ,
  \label{obs-w*-w}  \\
  \mathfrak{w}_{k m n} \mathfrak{w}^{*k m n } &=&
  \mathfrak{i}^k{}_k
  \mathfrak{i}^m{}_m
  \mathfrak{i}^n{}_n
    \, ,
  \ \ \ \ \mbox{\rm for different } k,m,n \ .
  \label{obs-w-w*}
\end{eqnarray}
We recall that
\begin{equation}
\label{1-j}
    \mathfrak{i}^k{}_k = {\bf 1} - \mathfrak{j}^k{}_k \, ,
\end{equation}
see \eqref{1-j1}. In the next subsection, we are going to prove
that the above properties uniquely characterize the algebra
${\mathfrak O}^{mat}_{\tree}$.

We show that the triality operator belongs to the center of the
algebra. For this purpose, note that $\mathfrak{j}^k{}_k $ is the
particle number operator at position $k \, .$ Thus,
\begin{equation}
\mathfrak{n} = \sum_{k=1}^{n}  \mathfrak{j}^k{}_k
\end{equation}
is the {\em total particle number operator}. By definition
\eqref{globalM} of ${\mathfrak{t}}_{\Lambda}$ we have
\begin{equation}
  \label{triality1} {\mathfrak{t}}_{\Lambda} = \varphi(\mathfrak{n})
  \ .
\end{equation}
This means that, in any representation, ${\mathfrak{t}}_{\Lambda}$
is equal to the particle number, modulo $3\, .$ As a direct
consequence of (\ref{commrel-j}), (\ref{commrel-jw}) and
\eqref{commrel-jw*}, we obtain
\begin{eqnarray}
\left[\mathfrak{n} \, ,\,  \mathfrak{j}^k{}_l \right] & = & 0 \,
\label{commrel-nj}\\
\left[\mathfrak{n} \, ,\, {\mathfrak{w}}^{*ijk} \right] & = & 3 \,
{\mathfrak{w}}^{*ijk} \, ,
\label{commrel-nw*} \\
\left[\mathfrak{n} \, ,\, \mathfrak{w}_{pqr} \right] & = & - 3 \,
\mathfrak{w}_{pqr} \, . \label{commrel-nw}
\end{eqnarray}
Together with \eqref{triality1}, these relations imply that all
generators and, thus, all hadronic observables commute with the
triality operator ${\mathfrak{t}}_{\Lambda}$.

\noindent
{\bf Remark:}\\
Due to \eqref{commrel-jw}, the whole set of baryonic invariants
$\mathfrak{w}_{l m n}$ can be generated from one chosen
$\mathfrak{w}_{{l_0} {m_0} {n_0}}$ by successively taking
commutators with $\mathfrak{j}`s$. Indeed, for $i \ne l$,
commutation relation \eqref{commrel-jw} reduces to the identity
$\left[ \mathfrak{j}^l{}_k , \mathfrak{w}_{l m n} \right]  = -
\mathfrak{w}_{k m n }$ which enables us to ``flip'' the
multi--index $(l,m,n)$ to any other position. All relations for
the remaining $\mathfrak{w}$`s then follow from the relations for
the single selected element $\mathfrak{w}_{{l_0} {m_0} {n_0}}$.


\subsection{Axiomatic Description}
\label{structure-j-w}


Consider now the abstract, unital $*$-algebra ${\cal A}$,
generated by abstract elements $\mathfrak{j}$, $\mathfrak{w}$ and
$\mathfrak{w}^*$, which fulfil relations \eqref{*-struct1} --
\eqref{obs-w-w*}. In the sequel, we shall prove that these
relations define the algebra uniquely, i.e. ${\cal A}$ is
identical with the previously defined algebra ${\mathfrak
O}^{mat}_{\tree}$. For this purpose, we derive from the defining
relations of ${\cal A}$ a number of additional identities.

\begin{Theorem}
The defining relations \eqref{*-struct1} -- \eqref{obs-w-w*} of
${\cal A}$ imply the following additional identities:
\begin{enumerate}
    \item
\begin{eqnarray}
\label{ll1}
\mathfrak{j}^m{}_l \mathfrak{j}^k{}_l  &=&  \delta^k{}_l
\, \mathfrak{j}^m{}_l \ , \\
\label{kk1} \mathfrak{j}^k{}_l \mathfrak{j}^k{}_n  &=&
\delta^k{}_l \, \mathfrak{j}^k{}_n\ .
\end{eqnarray}
Thus, in particular, the off-diagonal generators are nilpotent:
\begin{equation}
  \label{nilpotent1} \left(\mathfrak{j}^k{}_l \right)^2  =  0
  \ \ \ \ \ \mbox{\rm for} \ k \ne l \ .
\end{equation}
    \item
\begin{eqnarray}
 \label{jw1} \mathfrak{j}^i{}_k \, \mathfrak{w}_{lmn} &=& -
 \mathfrak{j}^i{}_l \, \mathfrak{w}_{kmn} = - \mathfrak{j}^i{}_m \,
 \mathfrak{w}_{lkn} = - \mathfrak{j}^i{}_n \mathfrak{w}_{lmk} \, .
 \\
 \label{jw*1} \mathfrak{w}^{*lmn}\ \mathfrak{j}^k{}_i  &=& -
 \mathfrak{w}^{*kmn}\ \mathfrak{j}^l{}_i   = -
 \mathfrak{w}^{*lkn}\ \mathfrak{j}^m{}_i
 = -  \mathfrak{w}^{*lmk} \mathfrak{j}^n{}_i .
\end{eqnarray}
In particular, $\mathfrak{j}^i{}_k$ multiplied with $\mathfrak{w}$
vanishes, if at least one of the indices of $\mathfrak{w}$
coincides with $k \, ,$ e.g.
\begin{equation}
\label{jw=0}
    \mathfrak{j}^k{}_l \mathfrak{w}_{l m n} = 0 \ , \ \ \ \
    \mathfrak{w}^{*l m n} \ \mathfrak{j}^l{}_k = 0 \ .
\end{equation}
    \item
The following identities hold;
\begin{eqnarray}
\label{corol--w}
  \mathfrak{j}^l{}_l \mathfrak{w}_{l m n} = &0& =
  \mathfrak{w}_{l m n} \mathfrak{i}^l{}_l \label{wi=0}\\
  \mathfrak{i}^l{}_l \mathfrak{w}_{l m n} = &\mathfrak{w}_{l m n}& =
  \mathfrak{w}_{l m n} \mathfrak{j}^l{}_l \label{iw=w}\\
  \mathfrak{w}^{*l m n} \mathfrak{j}^l{}_l = &0& =
  \mathfrak{i}^l{}_l \mathfrak{w}^{*l m n} \label{jw}\\
  \mathfrak{w}^{*l m n} \mathfrak{i}^l{}_l =&\mathfrak{w}^{*l m n}& =
  \mathfrak{i}^l{}_l \mathfrak{w}^{*l m n} \, . \label{iw}
\end{eqnarray}
  \item
The generators ${\mathfrak{w}}$ and ${\mathfrak{w}}^*$ are
nilpotent:
\begin{eqnarray}
\left( {\mathfrak{w}}^{*ijk} \right)^2 & = & 0 \, ,
\label{nilpotw*} \\
\left( {\mathfrak{w}}_{pqr} \right)^2 & = & 0 \, . \label{nilpotw}
\end{eqnarray}

\end{enumerate}

\end{Theorem}

\noindent {\bf Proof:}
\begin{enumerate}
\item To show relations \eqref{ll1}, we have to prove
\begin{eqnarray}
  \mathfrak{j}^m{}_l \mathfrak{j}^k{}_l  & = & 0 \, ,
  \quad \text{for} \quad k \neq l \ ,
  \label{ll}\\
  \mathfrak{j}^m{}_l \mathfrak{j}^l{}_l  & = & \mathfrak{j}^m{}_l   \ .
  \label{ml}
\end{eqnarray}
{}From \eqref{commrel-j} we get
\begin{equation}\label{pomoc}
\mathfrak{j}^k{}_l = \left[\mathfrak{j}^k{}_l ,
\mathfrak{j}^l{}_l \right] =
\mathfrak{j}^k{}_l \mathfrak{j}^l{}_l - \mathfrak{j}^l{}_l
\mathfrak{j}^k{}_l \, .
\end{equation}
Multiplying this relation by $\mathfrak{j}^l{}_l$ to the left and
using \eqref{idempotent1} yields
$$
\mathfrak{j}^l{}_l \mathfrak{j}^k{}_l =
\mathfrak{j}^l{}_l \mathfrak{j}^k{}_l \mathfrak{j}^l{}_l
- \mathfrak{j}^l{}_l \mathfrak{j}^k{}_l \, ,
$$
or, inserting expression \eqref{pomoc}  for $\mathfrak{j}^k{}_l$,
$$
2 \mathfrak{j}^l{}_l \mathfrak{j}^k{}_l=
\mathfrak{j}^l{}_l \mathfrak{j}^k{}_l \mathfrak{j}^l{}_l =
\mathfrak{j}^l{}_l \left(\mathfrak{j}^k{}_l \mathfrak{j}^l{}_l -
\mathfrak{j}^l{}_l \mathfrak{j}^k{}_l \right) \mathfrak{j}^l{}_l = 0 \, .
$$
This proves \eqref{ll} for the case $m =l \, .$ Consequently,
\eqref{pomoc} reduces to
\begin{equation}
\label{trick-left}
\mathfrak{j}^k{}_l = \mathfrak{j}^k{}_l \mathfrak{j}^l{}_l  \ .
\end{equation}
This proves relation \eqref{ml}. Finally, multiplying \eqref{ml}
by $\mathfrak{j}^k{}_l$ to the left and using $\mathfrak{j}^l{}_l
\mathfrak{j}^k{}_l = 0 $  yields \eqref{ll} for $m \neq l \, .$
The proof of relations \eqref{kk1} is completely analogous and,
therefore, we omit it here. 
\item We show the first relation in
\eqref{jw1},
$$
\mathfrak{j}^i{}_k \, \mathfrak{w}_{lmn} = - \mathfrak{j}^i{}_l \,
\mathfrak{w}_{kmn}\, .
$$
First, the commutation relations \eqref{commrel-jw} imply:
\begin{equation}
\label{pom}
  \mathfrak{j}^l{}_l \mathfrak{w}_{l m n} - \mathfrak{w}_{l m n}
  \mathfrak{j}^l{}_l = - \mathfrak{w}_{l m n} \ .
\end{equation}
Multiplying this equation from both sides by $\mathfrak{j}^l{}_l$
we obtain
$$
  \mathfrak{j}^l{}_l \mathfrak{w}_{l m n} \mathfrak{j}^l{}_l = 0 \ .
$$
Hence, multiplying \eqref{pom} from the left by
$\mathfrak{j}^l{}_l$ yields
$$
  \mathfrak{j}^l{}_l \mathfrak{w}_{l m n} =
  - \mathfrak{j}^l{}_l \mathfrak{w}_{l m n} \ ,
$$
or $\mathfrak{j}^l{}_l \mathfrak{w}_{l m n} =0$. Multiplying this
relation to the left by $\mathfrak{j}^k{}_l$ and using \eqref{ml}
yields
\begin{equation}
\label{jw=0-1}
    \mathfrak{j}^k{}_l \mathfrak{w}_{l m n} = 0 \ ,
\end{equation}
showing the special case \eqref{jw=0}. Now, multiplying the
commutation relations \eqref{commrel-jw} to the left by
$\mathfrak{j}^i{}_l$ and using \eqref{kk1} together with
\eqref{jw=0-1} gives $\delta^i{}_l \, \mathfrak{j}^i{}_k
\mathfrak{w}_{l m n }= - \delta^i{}_l \, \mathfrak{j}^i{}_l
\mathfrak{w}_{k m n }$ or, equivalently,
\begin{equation}
\label{jw=0-2}
\mathfrak{j}^l{}_k \mathfrak{w}_{l m n} = - \mathfrak{j}^l{}_l
\mathfrak{w}_{k m n} \ .
\end{equation}
Finally, multiplying this equation by $\mathfrak{j}^i{}_l$ and
using \eqref{commrel-jw}, \eqref{jw=0-1} and \eqref{ml} yields the
proof of the statement. The proof of the remaining equations
contained in \eqref{jw1} is identical.

\item First, observe that by \eqref{jw=0-1}, the auxiliary
identity \eqref{pom} reduces to
\begin{equation}
\label{pom1}
   - \mathfrak{w}_{l m n}
  \mathfrak{j}^l{}_l = - \mathfrak{w}_{l m n} \ ,
\end{equation}
and, hence, we have $\mathfrak{w}_{l m n} \mathfrak{i}^l{}_l =0$.
This way \eqref{wi=0} and \eqref{iw=w} are proved. Acting with the
operator $*$ on both sides we obtain the remaining identities
\eqref{jw} and \eqref{iw}.

\item Identity \eqref{wi=0} and \eqref{iw=w} imply nilpotency of
$\mathfrak{w}$:
$$
{\mathfrak{w}}_{lmn} {\mathfrak{w}}_{lmn} = {\mathfrak{w}}_{lmn}
\left(\mathfrak{i}^l{}_l {\mathfrak{w}}_{lmn} \right) =
\left({\mathfrak{w}}_{lmn} \mathfrak{i}^l{}_l \right)
{\mathfrak{w}}_{lmn}  = 0 \, .
$$
Similarly, nilpotency of $\mathfrak{w}^* $ follows from the
remaining two identities.

\end{enumerate}
\qed

Finally, observe that the idempotency and nilpotency properties
\eqref{idempotent1} and \eqref{nilpotent1} render ${\cal A}$
finite--dimensional. To summarize, ${\cal A}$ is a
(finite-dimensional) associative unital $*$-algebra. It is
obtained from the free algebra, generated by elements
$\{{\mathfrak{j}},{\mathfrak{w}},{\mathfrak{w}}^*, {\bf 1}\}$, by
factorizing with respect to the relations listed above.

For the sake of completeness, we have listed additional
interesting identities, see Appendix \ref{add--rel}, which have to
be taken into account, if one wants to build arbitrary monomials
in the generators.


\subsection{Irreducible Representations}
\label{irreps-rozdzial}


\begin{Lemma}
\label{CAR-irreps} There is at least one nontrivial, faithful
irreducible representation of ${\cal A}$, for each
eigenvalue $-1,0,1$ of the triality operator
${\mathfrak{t}}_{\Lambda}\, .$
\end{Lemma}
\noindent {\bf Proof:} Take the CAR--algebra $\cal C$ given by
\begin{equation}\label{a}
\left\{ {\mathfrak{a}}^*{}^k \, , \, {\mathfrak{a}}_k \, | \, \, k
= 1,2, \dots , n \right\} \, ,
\end{equation}
and fulfilling canonical anticommutation relations,
\begin{equation}
\label{a*a} \left[{\mathfrak{a}}^*{}^k \, , \, {\mathfrak{a}}_l
\right]_+ = \delta^k{}_l  \, .
\end{equation}
Denote its unique Hilbert representation space by $H$ and define
\begin{eqnarray}
\mathfrak{j}^k{}_l & = & {\mathfrak{a}}^*{}^k \, {\mathfrak{a}}_l
\, ,
\label{j-od-a}\\
\mathfrak{w}_{pqr} & = & {\mathfrak{a}}_p \, {\mathfrak{a}}_q \,
{\mathfrak{a}}_r \, ,
\label{w-od-a} \\
{\mathfrak{w}}^*{}^{ijk} & = & {\mathfrak{a}}^*{}^k \,
{\mathfrak{a}}^*{}^j \, {\mathfrak{a}}^*{}^i \,. \label{w*-od-a}
\end{eqnarray}
These abstract elements fulfil, of course, all relations of ${\cal
A}$ listed above. Thus, $H$ carries a representation of ${\cal
A}\, .$ Since ${\mathfrak{t}}_{\Lambda}$ commutes with all
$\mathfrak{j}$`s, $\mathfrak{w}$`s and  ${\mathfrak{w}}^*$`s, $H$
decomposes into superselection sectors,
\[
  H = H_{-1} \oplus H_0 \oplus H_1 \ ,
\]
corresponding to different eigenvalues of
${\mathfrak{t}}_{\Lambda}$. Each of these subspaces is invariant
under the action of ${\mathfrak{j}}$'s, ${\mathfrak{w}}$'s and
${\mathfrak{w}}^*$'s, providing a nontrivial, faithful and
irreducible representation of ${\cal A}$. \qed

\begin{Theorem}
\label{classif--irreps} Any irreducible, nontrivial representation
of ${\cal A}$ is equivalent to one of the three irreducible
representations provided by Lemma \ref{CAR-irreps}.
\end{Theorem}

For purposes of the proof, let us denote:
\begin{equation}
\label{E-projektory}
    E_{\nu_1 , \nu_2 \dots \nu_n}  :=
  \left( \mathfrak{j}^1{}_1 \right)^{\nu_1}
  \left( \mathfrak{i}^1{}_1 \right)^{\nu_1 +1}
  \left( \mathfrak{j}^2{}_2 \right)^{\nu_2}
  \left( \mathfrak{i}^2{}_2 \right)^{\nu_2 +1} \cdots
  \left( \mathfrak{j}^n{}_n \right)^{\nu_n}
  \left( \mathfrak{i}^n{}_n \right)^{\nu_n +1} \ ,
\end{equation}
where all indices $\nu_k$ assume values $0$ or $1$ and the
summation is meant {\em modulo} 2. Since the $\mathfrak{i}$`s and
$\mathfrak{j}$`s are Hermitean, orthogonal and commuting
projectors, $\{ E_{\nu_1 , \nu_2 \dots \nu_n}  \}$ is a family of
Hermitean orthogonal and commuting projectors, too. Moreover, we
have an obvious
\begin{Corollary}
The above projectors sum up to the unit element:
\begin{equation}
\label{rzuty}
    \bigoplus E_{\nu_1 , \nu_2 \dots \nu_n} = {\bf 1} \ .
\end{equation}
\end{Corollary}

\begin{Lemma}
\label{E--rel} The following relations hold for arbitrary $k \ne
l$:
\begin{enumerate}
\item $\mathfrak{j}^k{}_l \cdot E_{\nu_1 , \nu_2 \dots \nu_n} = 0$
unless $\nu_k = 0$  and $\nu_l = 1$. \item $E_{\nu_1 , \nu_2 \dots
\nu_n} \cdot \mathfrak{j}^k{}_l = 0$ unless $\nu_k = 1$ and $\nu_l
= 0$. \item For $\nu_k = 0$  and $\nu_l = 1$ we have
$\mathfrak{j}^k{}_l \cdot E_{\nu_1 , \dots , \nu_k , \dots , \nu_l
, \dots , \nu_n} = E_{\nu_1 , \dots , \nu_k +1 , \dots , \nu_l -1
, \dots , \nu_n} \cdot \mathfrak{j}^k{}_l$
\end{enumerate}
\end{Lemma}

\noindent {\bf Proof:} The proof follows by direct inspection from
the following identities, (which are all simple consequences of
\eqref{ll1} and  \eqref{kk1}):
$$
  \mathfrak{j}^k{}_l \mathfrak{j}^l{}_l \mathfrak{j}^k{}_k =
  0 \, ,\quad
  \mathfrak{j}^k{}_l \mathfrak{j}^l{}_l \mathfrak{i}^k{}_k =
  \mathfrak{j}^k{}_l \, , \quad
  \mathfrak{j}^k{}_l \mathfrak{i}^l{}_l \mathfrak{j}^k{}_k =
  0 \, , \quad
  \mathfrak{j}^k{}_l \mathfrak{i}^l{}_l \mathfrak{i}^k{}_k =
  0
$$
and
$$
   \mathfrak{j}^l{}_l \mathfrak{j}^k{}_k \mathfrak{j}^k{}_l =
  0 \, , \quad
   \mathfrak{j}^l{}_l \mathfrak{i}^k{}_k \mathfrak{j}^k{}_l =
  0 \, , \quad
   \mathfrak{i}^l{}_l \mathfrak{j}^k{}_k \mathfrak{j}^k{}_l =
  \mathfrak{j}^k{}_l  \, , \quad
   \mathfrak{i}^l{}_l \mathfrak{i}^k{}_k \mathfrak{j}^k{}_l =
  0 \, .
$$
\qed

\noindent {\bf Proof of the theorem:} Take such an irreducible
representation. Since the triality operator
$\mathfrak{t}_{\Lambda}$ lies in the center of ${\cal A}\, ,$ it
corresponds to a fixed value of triality. Take any other two
irreducible representations, corresponding to the remaining values
of triality. Let us denote these three representations by ${\cal
H}_t\, ,$ with $t = -1,0,1 \, .$ We are going to prove that there
exist isomorphisms
\begin{equation}
\label{equivalence}
    U_t : {\cal H}_t \rightarrow H_t
\end{equation}
intertwining the representations ${\cal H}_t$ with the three
CAR-representations $H_t \, ,$ defined in Lemma \ref{CAR-irreps}.
This will be accomplished by defining operators
$$
{\mathfrak{c}}^*{}^k : {\cal H}_t \rightarrow {\cal H}_{t+1}
$$
and
$$
{\mathfrak{c}}{}_k : {\cal H}_t \rightarrow {\cal H}_{t-1}
$$
(with summation {\em modulo} 3), fulfilling the CAR and such that
equations \eqref{j-od-a} -- \eqref{w*-od-a} are satisfied with
${\mathfrak{a}}$'s replaced by ${\mathfrak{c}}$'s. Then, the
statement of the theorem is a consequence of the classical
uniqueness theorem for CAR-representations.

Let us denote
\begin{equation}\label{obsadzenia}
    {\cal H}_{\nu_1 , \nu_2 \dots \nu_n} :=
    E_{\nu_1 , \nu_2 \dots \nu_n} {\cal H}_t \ .
\end{equation}
We obviously have
$$
\mathfrak{t}_{\Lambda} E_{\nu_1 , \nu_2 \dots \nu_n} =
E_{\nu_1 , \nu_2 \dots \nu_n} {\mathfrak{t}}_{\Lambda}
= t \, E_{\nu_1 , \nu_2 \dots \nu_n} \, .
$$
On the other hand, formula \eqref{triality1} implies
$$
\mathfrak{t}_{\Lambda} E_{\nu_1 , \nu_2 \dots \nu_n} =
\left( \sum_{i=1}^{n} \nu_i \ \mbox{\rm mod}\ 3 \right)
E_{\nu_1 , \nu_2 \dots \nu_n} \, .
$$
Thus, the only non-trivial subspaces are those fulfilling the
condition
\begin{equation}
\label{t=sum}
    t = \sum_{i=1}^{n} \nu_i \ \mbox{\rm mod}\ 3 \ .
\end{equation}
This fact, together with \eqref{rzuty}, implies
\begin{equation}\label{decomp}
    {\cal H}_t = \bigoplus_{t=\sum_{i=1}^{n} \nu_i \
    \mbox{\rm{\tiny mod}}\ 3}
    {\cal H}_{\nu_1 , \nu_2 \dots \nu_n} \ .
\end{equation}
Now, Lemma \ref{E--rel} implies that
$\mathfrak{j}^k{}_l {\cal H}_{\nu_1 ,\nu_2 \dots \nu_n} = 0 \, ,$
unless $\nu_k = 0$  and $\nu_l = 1$ and
that, in the latter case, $\mathfrak{j}^k{}_l$ maps
${\cal H}_{\nu_1 , \dots , \nu_k , \dots , \nu_l , \dots , \nu_n}$
onto
${\cal H}_{\nu_1 , \dots , \nu_k +1 , \dots , \nu_l -1 , \dots , \nu_n}$.
Observe, that $\mathfrak{j}^k{}_l$ is an isomorphism of
these two Hilbert spaces, with the inverse given by
$\mathfrak{j}^l{}_k$. Indeed, we have:
$ \mathfrak{j}^l{}_k \mathfrak{j}^k{}_l - \mathfrak{j}^k{}_l \mathfrak{j}^l{}_k =
\mathfrak{j}^l{}_l - \mathfrak{j}^k{}_k $ and, consequently,
$$
  \left( \mathfrak{j}^k{}_l \right)^* \mathfrak{j}^k{}_l =
  \mathfrak{j}^l{}_k \mathfrak{j}^k{}_l = \mathfrak{j}^l{}_l -
  \mathfrak{j}^k{}_k + \mathfrak{j}^k{}_l \mathfrak{j}^l{}_k \ .
$$
By Lemma \ref{E--rel}, this gives, for $\nu_k = 0$  and $\nu_l = 1$,
\begin{equation}
\label{izom}
    \left( \mathfrak{j}^k{}_l \right)^* \mathfrak{j}^k{}_l
    E_{\nu_1 , \dots , \nu_k , \dots , \nu_l , \dots , \nu_n}
    = E_{\nu_1 , \dots , \nu_k , \dots , \nu_l , \dots , \nu_n} \ .
\end{equation}
Similarly, relations \ref{corol--w} imply that $\mathfrak{w}_{lmn}
{\cal H}_{\dots , \nu_l ,\dots ,\nu_m ,\dots ,\nu_n,\dots} = 0 \,
,$ unless $\nu_l = \nu_m = \nu_n = 1$ and, in the latter case, it
maps ${\cal H}_{\dots , \nu_l ,\dots ,\nu_m ,\dots ,\nu_n,\dots}$
onto ${\cal H}_{\dots , \nu_l-1 ,\dots ,\nu_m-1 ,\dots
,\nu_n-1,\dots}$. Observe that, according to \eqref{obs-w*-w} and
\eqref{obs-w-w*}, $\mathfrak{w}_{lmn}$ is an isomorphism of these
two Hilbert spaces, with the inverse given by
$\mathfrak{w}^{*lmn}$.

Since the representations ${\cal H}_t$ are non-trivial, there is
at least one non-vanishing vector in at least one of the subspaces
${\cal H}_{\nu_1 , \nu_2 \dots \nu_n}\, ,$ for every ${\cal H}_t
\, .$ Thus, let us choose three such {\em normalized} vectors and
denote them by
$$
  |\nu_1(t) , \nu_2(t) \dots \nu_n(t) > \  \in {\cal
  H}_{\nu_1(t) , \nu_2(t) \dots \nu_n(t)} \ ,
$$
where $\sum_{i=1}^{n} \nu_i(t) \ \mbox{\rm mod}\ 3 =t$,
$t=-1,0,1.$ Acting with operators $\mathfrak{j}^k{}_l$,
$\mathfrak{w}_{lmn}$ and $\mathfrak{w}^{*lmn}$ on each of these
three vectors, we obtain, for every $t$,  a normalized vector, say
$|\nu_1 , \nu_2 \dots \nu_n >$, in each of the subspaces ${\cal
H}_{\dots , \nu_l ,\dots ,\nu_m ,\dots ,\nu_n,\dots}$. (The
information about $t$ is encoded implicitly, see equation
\eqref{t=sum}.) Moreover, we can label the vectors in the
representation spaces ${\cal H}_{\nu_1 , \nu_2 \dots \nu_n}$ in
such a way that the following relations are fulfilled:
\begin{eqnarray}
  \mathfrak{j}^k{}_l | .. , \nu_k , .. , \nu_l ,
  .. > &= & \left\{
  \begin{array}{l}
  \sigma(k,l)\cdot | .. , \nu_k+1 , .. , \nu_l-1 ,
  .. >  \\ \mbox{\rm if} \ \nu_k=0, \ \nu_l = 1 \ ,\\
  0 \ \ \mbox{\rm otherwise}\ ,
  \end{array}
  \right. \nonumber \\
  \mathfrak{w}_{lmn} | .., \nu_l , .. , \nu_m , .. , \nu_n , .. >
  &= & \left\{
  \begin{array}{l}
  \sigma(l,m,n)\cdot
  | .., \nu_l-1 , .. , \nu_m-1 , .. , \nu_n-1 , ..>
   \\ \mbox{\rm if} \ \nu_l=\nu_m = \nu_n=1 \ , \\
  0 \ \ \mbox{\rm otherwise}\ ,
  \end{array}
  \right. \nonumber \\
  \mathfrak{w}^{*lmn} | .. , \nu_l , ..  , \nu_m , .. , \nu_n ,.. >
  &= & \left\{
  \begin{array}{l}
  - \sigma(l,m,n)\cdot
  | .. , \nu_l+1 , .. , \nu_m+1 , .. , \nu_n+1 ,..  >
   \\ \mbox{\rm if} \ \nu_l=\nu_m = \nu_n=0 \ , \\
  0 \ \ \mbox{\rm otherwise} \ ,
  \end{array}
  \right. \nonumber
\end{eqnarray}
where
\begin{eqnarray}
  \sigma(k,l) &=& (-1)^{s(k)-s(l)} \ , \nonumber\\
  s(k) &=& \sum_{i < k} \nu_i \ , \nonumber\\
  \sigma(l,m,n) &=& s(l,m,n)(-1)^{s(l)+s(m)+s(n)} \ ,
  \nonumber
\end{eqnarray}
and $s(l,m,n)$ is the sign of the permutation which is necessary
to sort the triple $(l,m,n)$ in growing order (i.e.
$s(l,m,n)=1$ if $l < m < n$, $s(l,m,n)=-1$ if $l < n < m$ etc.).

We show that this labelling is possible, indeed: We start with
three arbitrarily chosen vectors given by sequences of $\nu_i =
\nu_i(t)$, for $t=-1,0,1 \, .$ Next, we apply operators
$\mathfrak{j}^k{}_l$, $\mathfrak{w}_{lmn}$ and
$\mathfrak{w}^{*lmn}$ to these vectors and use the above formulae
as the definition of the corresponding vectors on the right hand
side. Now, it remains to prove that this definition does not
depend upon the order of these operations. For this purpose, we
use the commutation rules \eqref{commrel-j} and
\eqref{commrel-jw}. As far as the commutation relations
$[\mathfrak{w},\mathfrak{w}]$, $[\mathfrak{w}^*,\mathfrak{w}]$ and
$[\mathfrak{w}^*,\mathfrak{w}^*]$ are concerned, we can use
relations \eqref{jw1} -- \eqref{jw=0}  to flip the indices of
occuring $\mathfrak{w}$`s and $\mathfrak{w}^*$` in such a way
that, whenever these objects meet, they have always the same
indices. Then, we use relations \eqref{obs-w*-w} and
\eqref{obs-w-w*} together with nilpotency properties
\eqref{nilpotw*} and \eqref{nilpotw}. Having done this, the
formula may be checked by inspection.

Because of the irreducibility of the representations ${\cal H}_t$,
the vectors $|\nu_1(t) , \nu_2(t) \dots \nu_n(t) >$ form
(orthogonal) bases in each ${\cal H}_t$. Hence, we define the
intertwining operator $U$ putting:
\[
 U |\nu_1 , \nu_2 \dots \nu_n > :=  \left( {\mathfrak{a}}^*{}^1
 \right)^{\nu_1} \left( {\mathfrak{a}}^*{}^2
 \right)^{\nu_2} \cdots \left( {\mathfrak{a}}^*{}^n
 \right)^{\nu_n}|0> \ ,
\]
where ${\mathfrak{a}}^*$'s are the CAR-creation operators from
Lemma \ref{CAR-irreps} and $|0>\ \in H$ is the Fock vacuum. (The
label $t$ has been ommited.) Then, the operators
$$
  {\mathfrak{c }}^* := U^{-1} {\mathfrak{a}}^* U
$$
and
\[
  {\mathfrak{c }} := U^{-1} {\mathfrak{a}} U
\]
satisfy the CAR. It is easy to check that they fulfill equations
\eqref{j-od-a} -- \eqref{w*-od-a}, with ${\mathfrak{a}}$ replaced
by ${\mathfrak{c}}$. This ends the proof.
\qed

This theorem shows that any algebra ${\cal A}$ generated by
abstract elements $\mathfrak{j}$, $\mathfrak{w}$ and
$\mathfrak{w}^*$, fulfilling relations \eqref{*-struct1} --
\eqref{obs-w-w*}, is isomorphic to the commutant of
the triality operator
$$
\mathfrak{t} = \varphi\left(\sum_k {\mathfrak{a}}^{*k}{\mathfrak{a}}_k\right)
$$
in ${\cal C}$,
\begin{equation}
\label{alg--iso}
\cal A \cong \mathfrak{t}^{\prime}({\cal C}) \subset \cal C \,.
\end{equation}
This implies the following
\begin{Corollary}
The algebras ${\cal A}$ and ${\mathfrak O}^{mat}_{\tree}$ are isomorphic.
\end{Corollary}


\subsection{Super Lie Structure}
\label{Super Lie Structure}


Formula \eqref{alg--iso} provides us with a simple and nice
algebraic characterization of ${\mathfrak O}^{mat}_{\tree}\, .$
Nonetheless, since in the case of lattice QED, we have found a Lie
algebraic characterization of the matter field part \cite{KRS,
KRS1}, it is worthwile to ask, whether a similar characterization
is possible in QCD as well. The answer is affirmative, as we show
now.

Using an idea of Palev \cite{Palev}, see also Dondi and Jarvis \cite{Jarvis}, 
we define the following operators:
\begin{eqnarray}
{\mathfrak{b}}^*{}^k & := & {\phi}^*{}^k \, \sqrt{p - \mathfrak{n}} \, ,
\label{b^*k} \\
\mathfrak{b}_k  & := & \sqrt{p - \mathfrak{n}} \, {\phi}_k \, ,
\label{b^k}
\end{eqnarray}
with $p$ being a positive integer. In what follows we use the
following obvious formulae:
\begin{eqnarray}
\label{comm-af}
{\phi}_k \, f(\mathfrak{n}) & = & f(\mathfrak{n} + 1)
\, {\phi}_k \, , \\
\label{comm-a*f}
{\phi}^{*k} \, f(\mathfrak{n}) & = & f(\mathfrak{n} - 1)
\, {\phi}^{*k} \, ,
\end{eqnarray}
for any operator function $f\,.$ In terms of the
$\mathfrak{b}$-operators, the (anti-)commutation  relations take
the following form:
\begin{eqnarray}
\label{commrel-j1}
  \left[ \mathfrak{j}^k{}_l , \mathfrak{j}^m{}_n \right] & = &
  \delta^m{}_l \, \mathfrak{j}^k{}_n -
  \delta^k{}_n \, \mathfrak{j}^m{}_l \, , \\
\left[ \mathfrak{j}^k{}_l, {\mathfrak{b}}^*{}^i \right] & = &
\delta^i{}_l \,{\mathfrak{b}}^*{}^k \, ,
\label{commrel-j*b}\\
\left[ \mathfrak{j}^k{}_l, \mathfrak{b}_i \right] & = &
- \delta^k{}_i \, \mathfrak{b}_l \, ,
\label{commrel-jb}\\
\left[{\mathfrak{b}}^*{}^k \, , \, {\mathfrak{b}}_l \right]_+
& = &(p - \mathfrak{n}) \, \delta^k{}_l
+  \mathfrak{j}^k{}_l \, .
\label{gaugeinvfermions-b}
\end{eqnarray}
This shows that
\begin{equation}
\label{linenv}
\mathfrak{A} := lin.env.\left\{{\mathfrak{b}}^*{}^k,
{\mathfrak{b}}_k, \mathfrak{j}^k{}_l \, \,
| \, \,  k,l = 1,2, \dots ,n  \right\}
\end{equation}
is isomorphic to the Lie superalgebra ${\rm sl(1/n)} \, .$ In more
detail, identifying
\begin{equation}
e^k{}_l  =  \mathfrak{j}^k{}_l - \tfrac{1}{N} \, \delta^k{}_l \,
\mathfrak{n} \, \, , \,\,
e^0{}_0  =   \tfrac{N}{N-1} p - \mathfrak{n} \, \, ,  \, \,
e^k{}_0  =  {\mathfrak{b}}^*{}^k \, \, , \, \,
e^0{}_k  =  \mathfrak{b}_k \, \, ,
\end{equation}
we obtain the standard (anti-)commutation relations for ${\rm sl(1/n)}$:
\begin{eqnarray}
\left[ e^k{}_l , e^m{}_n \right] & = & \delta^m{}_l \, e^k{}_n -
  \delta^k{}_n \, e^m{}_l \, , \nonumber\\
\left[ e^0{}_0, e^i{}_0 \right] & = & - e^i{}_0  \, ,
\nonumber \\
\left[ e^0{}_0, e^0{}_i \right] & = &  e^0{}_i  \, ,
\nonumber \\
\left[ e^k{}_l, e^i{}_0 \right] & = &
\delta^i{}_l \, e^k{}_0 - \tfrac{1}{N} \,  \delta^k{}_l \, e^i{}_0 \, ,
\nonumber \\
\left[ e^k{}_l, e^0{}_i \right] & = &
- \delta^k{}_i \, e^0{}_l + \tfrac{1}{N} \, \delta^k{}_l \, e^0{}_i  \, ,
\nonumber \\
\left[e^k{}_0 \, , \, e^0{}_l \right]_+ & = & e^k{}_l + \tfrac{N-1}{N} \,
\delta^k{}_l e^0{}_0 \, .
\nonumber
\end{eqnarray}
The even part is isomorphic to ${\rm gl(n,\mathbb{C})}$,
$$
{\rm sl(1/n)}_{\bar 0} = {\rm gl(n,\mathbb{C})} =
lin.env.\left\{ e^k{}_l\, , \, e^0{}_0 \, \,  | \, \, k,l = 1,2,
\dots ,n \right\} \, ,
$$
and the odd part is given by
$$
{\rm sl(1/n)}_{\bar 1} = lin.env.\left\{e^k{}_0\, ,
\, e^0{}_k | \, \, k = 1,2, \dots ,n \right\} \,.
$$
Next, observe that
\begin{equation}
\label{bbb}
\mathfrak{b}_i \, \mathfrak{b}_j \, \mathfrak{b}_k =
\sqrt{F(\mathfrak{n})} \, \mathfrak{w}_{i j k} \, ,
\end{equation}
with
$$
F(\mathfrak{n}) = (p - \mathfrak{n}) \,(p -1  -
\mathfrak{n})\, (p - 2 - \mathfrak{n}) \, .
$$
{}From now on we assume
$$
p = n + 3 \,.
$$
Then $F(\mathfrak{n})$ is a positive operator in every
representation. Thus, in every representation we can express the
baryonic invariants $\mathfrak{w}$ in terms of the fermionic
operators $\mathfrak{b}$:
\begin{equation}
\label{w(bbb)}
\mathfrak{w}_{i j k} = F(\mathfrak{n})^{-\tfrac{1}{2}}\, \mathfrak{b}_i \,
\mathfrak{b}_j \, \mathfrak{b}_k \, .
\end{equation}
We denote
\begin{equation}
\label{tildew}
{\tilde {\mathfrak{w}}}_{i j k} =  \mathfrak{b}_i \,
\mathfrak{b}_j \, \mathfrak{b}_k \, ,
\end{equation}
and
\begin{equation}
\label{tildew*}
{\tilde {\mathfrak{w}}}^{*i j k} =  \mathfrak{b}^{*k} \,
\mathfrak{b}^{*j} \, \mathfrak{b}^{*i} \, .
\end{equation}
For the bosonic part, we implement relations \eqref{*-struct1} and
\eqref{idempotent1}
\begin{eqnarray}
 \left(\mathfrak{j}^k{}_l \right)^* & = & \mathfrak{j}^l{}_k \ ,
  \label{*-struct2}\\
\left(\mathfrak{j}^k{}_k \right)^2  & = & \mathfrak{j}^k{}_k \ .
\label{idempotent2}
\end{eqnarray}
see Subsection \ref{gen--rel}.
These relations define a Lie ideal $\mathfrak{I}$ in the
enveloping algebra
$$
\mathfrak{U}\left({\rm gl(n,\mathbb{C})}\right) \subset
\mathfrak{U}\left({\rm sl(1/n)}\right)\, ,
$$
by which we factorize. Moreover, we implement that every
observable has to commute with the triality operator.  Thus, we
have to take the commutant of $\mathfrak{t}$ in this factor
algebra, which we denote by
\begin{equation}
\cal L := \mathfrak{t}^{\prime}\left( \mathfrak{U}({\rm sl(1/n)})
/ \mathfrak{I} \right)\, .
\end{equation}

\begin{Theorem}
The associative unital $*$-algebras ${\cal A}$ and ${\cal L}$ are
isomorphic.
\end{Theorem}

\noindent
{\bf Proof:}\\
First, observe that the operations of taking the commutant and of
factorizing with respect to $\mathfrak{I}$ commute, because
$\mathfrak{t}$ commutes with every $\mathfrak{j}^k{}_k \ .$

To prove the above isomorphism, we show that ${\cal A}$ and ${\cal
L}$ have exactly the same irreducible representations. For that
purpose, recall that ${\rm sl(1/n)}$ is a basic Lie superalgebra
of type I, which means
\begin{equation}
\label{typeI}
{\rm sl(1/n)}_{\bar 1} = {\rm sl(1/n)}_{-1} \, \oplus \,
{\rm sl(1/n)}_{+1} \, ,
\end{equation}
with ${\rm sl(1/n)}_{-1}$ and ${\rm sl(1/n)}_{+1}$ being two
irreducible moduls of ${\rm sl(1/n)}_{\bar 0} \cong {\rm
gl(n,\mathbb{C})}$, in terms of our generators spanned by
$\left\{\mathfrak{b}_k \right\}$  and  $\left\{\mathfrak{b}^{*k}
\right\}$ respectively. It follows from general representation
theory, see \cite{Kac}, that any finite dimensional irreducible
representation of a basic Lie superalgebra $\mathfrak{G}$ is
obtained from a Kac module. For superalgebras of type I, every Kac
module $V(\lambda)$ is induced from a highest weight module
$V_0(\lambda)$ of the even part $\mathfrak{G}_{\bar 0}$:
\begin{equation}
\label{indmodule}
V(\lambda) = {\rm Ind}_{\mathfrak{K}}^{\mathfrak{G}} \, V_0(\lambda)
:= \mathfrak{U} \left( \mathfrak{G} \right) \otimes_{\mathfrak{U}
\left(\mathfrak{K} \right)} \, V_0(\lambda) \, ,
\end{equation}
where
\begin{equation}
\mathfrak{K} = \mathfrak{G}_{\bar 0} \, \oplus \, \mathfrak{G}_1
\end{equation}
and $\mathfrak{U} \left( \mathfrak{G} \right)$ and $\mathfrak{U}
\left( \mathfrak{K} \right)$ denote the enveloping algebras of
$\mathfrak{G}$ and $\mathfrak{K}$ respectively. Formula
(\ref{indmodule}) has to be understood as follows: The
$\mathfrak{G}_{\bar 0}$-module $V_0(\Lambda)$ has been extended to
a $\mathfrak{K}$-module by putting
$$
\mathfrak{G}_1 \,  V_0(\lambda) = 0
$$
and one has to identify elements
$$
k \otimes v = 1 \otimes k(v) \, ,
$$
for $k \in \mathfrak{K}$ and $v \in V_0(\lambda)\, .$ Then the
induced representation of $\mathfrak{G}$ is defined by
\begin{equation}
\label{indrep}
g(u \otimes v) := g\, u \, \otimes \, v \, ,
\end{equation}
for $g \in \mathfrak{G} \, ,$  $u \in \mathfrak{U}
\left(\mathfrak{G} \right)$ and $v \in V_0(\lambda)\, .$ We stress
that $V(\lambda)$ is not always simple. In that case, one has to
factorize by a certain maximal submodule, to obtain an irreducible
representation.

Now, let $V_0(\lambda)$ be a highest weight module of $
\mathfrak{G}_{\bar 0} = {\rm gl(n,\mathbb{C})}\, .$ Since in our
case $\left[\mathfrak{G}_{-1} \, , \, \mathfrak{G}_{-1}  \right]_+
= 0 \, ,$ we have
\begin{equation}
\label{indmodule1}
V(\lambda) \cong \Lambda \left( \mathfrak{G}_{-1} \right) \, \otimes \,
V_0(\lambda) \, ,
\end{equation}
with
$$
\Lambda \left( \mathfrak{G}_{-1} \right) =
\bigoplus_{k = 0}^n \Lambda^k \left( \mathfrak{G}_{-1} \right)
$$
denoting the exterior algebra of $\mathfrak{G}_1\ .$ Thus, in terms of
generators, we have
\begin{equation}
\label{indmodule2}
V(\lambda) \cong \bigoplus_{1<k_1< \dots <k_n \leq n} \, \mathfrak{b}_{k_1}
\dots \mathfrak{b}_{k_n} \, V_0(\lambda) \, .
\end{equation}
We show that taking the above commutant and factorizing with
respect to $\mathfrak{I}$ reduces the set of irreducible
representations to three inequivalent representations labelled by
triality.

First, in the commutant of $\mathfrak{t}$, only monomials in
$\mathfrak{b}$ and $\mathfrak{b}^*$ built from ${\tilde
{\mathfrak{w}}}$ and ${\tilde {\mathfrak{w}}}^*$ can occure. Thus,
$V(\lambda)$ takes the form:
\begin{equation}
\label{indmodule3}
V(\lambda) \cong \bigoplus_{1<i_1<j_1<k_1< \dots <i_n<j_n<k_n \leq N} \,
{\tilde {\mathfrak{w}}}_{i_1 j_1 k_1}  \dots
{\tilde {\mathfrak{w}}}_{i_n j_n k_n}
\, V_0(\lambda) \, .
\end{equation}
Since the $\tilde {\mathfrak{w}}$`s act transitively on this
direct sum, $V(\lambda)$ is an irreducible module. Moreover, as a
direct consequence of the commutation relations we have
\begin{eqnarray}
\left[\mathfrak{n} \, ,\,  \mathfrak{j}^k{}_l \right] & = & 0 \, ,
\label{commrel-nj1}\\
 \left[\mathfrak{n} \, ,\, {\tilde {\mathfrak{w}}}^{*ijk} \right]
 & = & 3 \, {\tilde {\mathfrak{w}}}^{*ijk} \, ,
\label{commrel-nw*1} \\
 \left[\mathfrak{n} \, ,\, {\tilde {\mathfrak{w}}}_{pqr} \right]
 & = & - 3 \, {\tilde {\mathfrak{w}}}_{pqr} \, .
\label{commrel-nw2}
\end{eqnarray}
Thus, in any representation, ${\tilde {\mathfrak{w}}}_{pqr}$
lowers the particle number by $3$, whereas ${\tilde
{\mathfrak{w}}}^{*ijk}$ raises it by $3$.

Next, by \eqref{idempotent2} the particle number operator
$\mathfrak{j}^k{}_k$ at position $k$ can take only eigenvalues $0$
and $1$, on any highest weight module $V_0(\lambda)$ of ${\rm
gl(n,\mathbb{C})}\, .$ Every highest weight module of ${\rm
gl(n,\mathbb{C})}$ is built -- by taking tensor products -- from
fundamental representations, which in turn are all isomorphic to
some exterior product $\Lambda^l (\mathbb{C}^n) \, .$ But,
whenever we take a tensor product of such exterior products, which
is not antisymmetric, there exists a vector, for which
$\mathfrak{j}^k{}_k$ has an eigenvalues greater than one. Thus,
\eqref{idempotent2} reduces the admissible highest weight modules to
the fundamental ones. Since the operators $\tilde {\mathfrak{w}}$
lower the particle number by $3$, the lowest weight component of
(\ref{indmodule3}) can have particle numbers $0,1$ or $2$, only.
Using the canonical basis of $\mathbb{C}^n$, an explicit
isomorphism intertwining these $3$ representations with the
representations $H_t$ can be written down, as in the proof of
Theorem \ref{classif--irreps}. \qed

\setcounter{equation}{0}
\section{Discussion}


\begin{enumerate}
\item
The generators $J^{a b}_{\gamma}(x,y)$ and  $W^{a b c}_{\alpha \beta
\gamma}(x,y,z)$ (see \eqref{observableJ} and \eqref{observableW})
are difficult to handle. This is why we have replaced
them by generators $\mathfrak{j}^k{}_l$ and $\mathfrak{w}_{pqr}$
(see \eqref{observ-a1} and \eqref{observ-a2}), fulfilling much
simpler relations. To define these observables  we have 
used a gauge fixing procedure based upon the choice of a tree. However, it is obvious that
the specific gauge we have chosen is irrelevant for the structure
of the algebra, defined by relations \eqref{*-struct1} --
\eqref{obs-w-w*}.  Changing the gauge condition does
not affect these relations. Thus, there should exist another, more intrinsic procedure 
for obtaining this algebra, which does not rely on gauge fixing.

To make this transparent,  assume that we have chosen a tree. Now, instead of fixing the gauge, we rewrite 
the operators $J$ and $W$ in terms of fermionic  operators parallel-transported to the tree root $x_0$. Denoting these transported operators by $\tilde{\psi}$, we get: 
\begin{equation}
\label{J--sigma}
J^{ab}_{\gamma}(x,y) = \tilde \psi^{*a}_A(x_0) U_{\sigma}^A{}_B \tilde \psi^{bB} (x_0) \ , 
\end{equation}
with
$$
\sigma = \beta \circ \gamma \circ \alpha^{-1}
$$
being the closed path uniquely defined by this parallel transport, ($\alpha$ and $\beta$ are the unique 
on-tree paths from $x$ resp. $y$ to $x_0 \, .$) Thus, the operators $J$ acquire a labelling by 
(unparameterized) closed paths.
 Collecting the spinorial index and the point $x\in
\Lambda$ into a single index $u = (a,x)$ , we get a mapping 
$\sigma \mapsto  J^u{}_v (\sigma)$.
It can be easily checked that
the commutation relations for the quantities $J$ then take the following form:
\begin{equation}\label{comm-J}
    \left[ J^u{}_v (\beta) , J^w{}_t(\gamma) \right] = \delta^w{}_v J^u{}_t (\beta \circ \gamma) -
    \delta^u{}_t J^w{}_v (\gamma  \circ \beta) \ ,
\end{equation}
where $\circ$ denotes the natural multiplication in the group of
(unparameterized) closed  paths. Similarly, the baryonic operators $W$ and $W^*$ can be rewritten , acquiring 
a labelling by  closed paths, $ \sigma \mapsto W^{uvw}( \sigma)$,
$ \sigma \mapsto W^*{}^{uvw}( \sigma)$. The anticommutation relations for $W$ and $W^*$
can be easily worked out, but we omit them here.
\item
Now, let us restrict ourselves to on--tree paths in $J$ and $W$ only. It is clear from \eqref{J--sigma} that for them 
all on-tree parallel transporters $U_{\sigma}$ are
equal to ${\bf 1}$,
yielding quantities $J^u{}_v $, $ W_{uvw}$,
$ W^*{}^{uvw}$. Thus the algebra labelled by closed paths descends to an algebra
defined by the following (anti-)commutation relations: 
\begin{eqnarray}
\label{comm-J1}
    \left[ J^u{}_v  , J^w{}_t \right] &=&
    \delta^w{}_v J^u{}_t -
    \delta^u{}_t J^w{}_v  \ , \\
 	{\{} W^*{}^{(uvw)}, W_{(pqr)} {\}} &=&
	  \frac 12{(J\! \cdot\! J\! \cdot \!\delta)^{(uvw)}}_{(pqr)} \!+\!2{(J \! \cdot \! 
	\delta \! \cdot \! \delta)^{(uvw)}}_{(pqr)} 
	\!-\!6{\delta^{(uvw)}}_{(pqr)} \, , \label{comm-WW*} \\
        {\{} W^*{}^{(uvw)}, W^*{}^{(pqr)} {\}} &=& 0, \quad {\{} W_{(uvw)}, W_{(pqr)} 
	{\}} = 0 \, , \label{comm-WW} \\
	{[}{J^{s}}_{t}, {W}^*\mbox{}^{(uvw)}{]} &=& 
	{\delta_{t}}^{u}{W}^*\mbox{}^{(s v w)} + 
	{\delta_{t}}^{v}{W}^*\mbox{}^{(u s w)} + 
	{\delta_{t}}^{w}{W}^*\mbox{}^{(u v s)} \, , \label{comm-JW*} \\
	{[}{J^{s}}_{t}, {W}_{(pqr)}{]} &=& 
	-{\delta^{s}}_{p} {W}_{(tqr)} -{\delta^{s}}_{q}{W}_{(ptr)} 
    -{\delta^{s}}_{r}{W}_{(pqt)} \, , \label{comm-JW} 	
\end{eqnarray}
Here  $(J \! \cdot \! J \! \cdot \! \delta)$ and $(J \! \cdot \! \delta \! \cdot \! \delta)$ 
are the appropriate  totally symmetric combinations. Allowing for cyclic permutations on $uvw$ and $pqr$, $(J \! \cdot \! J \! \cdot \!
\delta)$ contains a total of $4\times 9$ $=36$ terms, $(J \! \cdot \! \delta
\! \cdot \! \delta)$ contains $9 \times 2 = 18$ terms, and $(\delta \! \cdot \!
\delta \! \cdot \! \delta)$ just $3 \times 2 = 6$ terms:
\begin{eqnarray}
	{{(}J \! \cdot \! J \! \cdot \! \delta{)}^{(uvw)}}_{(pqr)} &=& 
	({J^{u}}_{p}{J^{v}}_{q} + {J^{v}}_{p}{J^{u}}_{q} + 
	           {J^{u}}_{q}{J^{v}}_{p} + {J^{v}}_{q}{J^{u}}_{p}){\delta^{w}}_{r} + \ldots , \nonumber \\
    {{(}J \! \cdot \! \delta \! \cdot \! \delta{)}^{(uvw)}}_{(pqr)} &=& 
	{J^{u}}_{p}({\delta^{v}}_{q}{\delta^{w}}_{r} + {\delta^{v}}_{r}{\delta^{w}}_{q}) + \ldots , \nonumber \\
    {\delta^{(uvw)}}_{(pqr)} &=& {\delta^{u}}_{p} ({\delta^{v}}_{q} 
    {\delta^{w}}_{r} + {\delta^{v}}_{r}{\delta^{w}}_{q})+ \ldots.
   \label{eq:SymmetricObjects}
\end{eqnarray}

Obviously, equations \eqref{comm-J1} are the commutation relations of $\rm gl(4N,\mathbb{C})$.
The anticommutator \eqref{comm-WW*}
closes on a quadratic polynomial 
in the enveloping algebra of the even (Lie) subalgebra $\rm gl(4N,\mathbb{C})$. 
Thus we have identified the $W$ and $W^*$
as odd generators of a 
supersymmetry algebra belonging to a class of `polynomial'  superalgebras. 
Such `nonlinear' extensions of Lie algebras and superalgebras have been recognised 
in other contexts in recent literature. An initial investigation of them in the case of 
generalisations of  gl(4N/1) (or more generally of type I Lie superalgebras) has been 
given in \cite{JR} (see also the related remarks in the appendix).

We stress that, again by formula \eqref{J--sigma}, the full set of operators $J$ and $W$ can be reconstructed, 
knowing the generators $J^u{}_v $, $ W^{uvw}$,
$ W^*{}^{uvw}$, together with the Wilson loops $U_{\sigma} \, .$

\item
Clearly, the quantities $J^u{}_v $, $ W_{uvw}$,
$ W^*{}^{uvw}$ constructed under point 2 can be viewed as obtained from on-tree 
gauge fixing (putting the parallel transporter on every on-tree link equal to $\bf 1$).
If we remove the residual gauge freedom (at the root), we can pass to the quantities
$\mathfrak{j}^k{}_l $, $\mathfrak{w}_{ijk}$, ${\mathfrak{w}}^*{}^{pqr}$ used in this paper.
In the case of a generic orbit we
have proved (see \cite{KR1}) that the representation $J^u{}_v (\beta)$ is
``sufficiently non-degenerate'', and we may reduce it to
$\mathfrak{j}^k{}_l$. Actually, this non-degeneracy follows from
the non-degeneracy of the representation of the electric fluxes
$E_{\gamma}(x,y)$ -- see \eqref{observableE}. We expect that
there exist  ``degenerate'' representations, related to
non--generic orbits, which do not allow to extract the
representation of the full $\rm gl(n,\mathbb{C})$ Lie algebra. Indeed, the impossibility to
fix the gauge completely on a non--generic orbit (having
a non--trivial stabilizer) implies the impossibility of
reconstructing the quantities $\mathfrak{j}^k{}_l$, because they
are not invariant with respect to the stabilizer. In this case, we
expect that the fermions ${\mathfrak{a}}_k$ carrying the
representation (see formulae \eqref{a} -- \eqref{w-od-a}) will be
replaced by some ``anyons'', satisfying (possibly) a different
statistics. A consistent mathematical analysis of such
representations of the observable algebra (if they do exist)
together with their physical implications will be one of our next goals.
\end{enumerate}
\appendix
\renewcommand{\theequation}{\Alph{section}.\arabic{equation}}


\setcounter{equation}{0}
\section{Additional Relations}
\label{add--rel}

Here, we list additional relations, also following from relations
\eqref{*-struct1} -- \eqref{obs-w-w*}.

First, we have the following so-called characteristic
identities:
\begin{equation}
\label{chid}
\mathfrak{j}^k{}_l \, \mathfrak{j}^l{}_m  =
(n + 1 - \mathfrak{n}) \mathfrak{j}^k{}_m \, ,
\end{equation}
(with the sum taken over all $l \, .$) Next, one can analyze
arbitrary higher order monomials, built from $\mathfrak{w}$ and
$\mathfrak{w}^*\, .$ For that purpose, let us introduce the
following tensor operators (totally antisymmetric in both upper
and lower indices) built from $\mathfrak{j}$`s:
\begin{eqnarray}
X^{i_1 i_2 i_3}{}_{p_1 p_2 p_3} & = &
\sum_{\rho,\sigma} {sgn(\rho)} sgn(\sigma) \,
\mathfrak{j}^{i_{\rho_1}}{}_{p_{\sigma_1}} \,
\mathfrak{j}^{i_{\rho_2}}{}_{p_{\sigma_2}}
\, \mathfrak{j}^{i_{\rho_3}}{}_{p_{\sigma_3}} \, ,
\label{X} \\
Y^{i_1 i_2 i_3}{}_{p_1 p_2 p_3} & = &\sum_{\rho,\sigma} sgn(\rho)
sgn(\sigma) \, \mathfrak{j}^{i_{\rho_1}}{}_{p_{\sigma_1}} \,
\mathfrak{j}^{i_{\rho_2}}{}_{p_{\sigma_2}} \,
\delta^{i_{\rho_3}}{}_{p_{\sigma_3}}  \, ,
\label{Y} \\
Z^{i_1 i_2 i_3}{}_{p_1 p_2 p_3} & = &
\sum_{\rho,\sigma} sgn(\rho) sgn(\sigma) \,
\mathfrak{j}^{i_{\rho_1}}{}_{p_{\sigma_1}}
\, \delta^{i_{\rho_2}}{}_{p_{\sigma_2}}
\, \delta^{i_{\rho_3}}{}_{p_{\sigma_3}}  \, .
\label{Z} \\
D^{i_1 i_2 i_3}{}_{p_1 p_2 p_3} & = &
\sum_{\rho,\sigma} sgn(\rho) sgn(\sigma) \,
\delta^{i_{\rho_1}}{}_{p_{\sigma_1}} \,
\delta^{i_{\rho_2}}{}_{p_{\sigma_2}}
\, \delta^{i_{\rho_3}}{}_{p_{\sigma_3}}  \, ,
\label{D}
\end{eqnarray}
with sums running over all permutations $\rho$ and $\sigma$.
Using \eqref{observ-a2} and \eqref{observ-a3}, a lengthy but
simple calculation yields:
\begin{eqnarray}
36 \, {\mathfrak{w}}^*{}^{ijk} \, \mathfrak{w}_{pqr} & = & X^{i j
k}{}_{p q r} + 3 Y^{i j k}{}_{p q r} + 2 Z^{i j k}{}_{p q r}  \, ,
\label{w*w} \\
36 \, \mathfrak{w}_{pqr} \, {\mathfrak{w}}^*{}^{ijk} & = &
- X^{i j k}{}_{p q r} + 6 Y^{i j k}{}_{p q r} - 11 Z^{i j k}{}_{p q r}
+ 6 D^{i j k}{}_{p q r} \, .
\label{ww*}
\end{eqnarray}
Using these relations and keeping in mind the nilpotency
properties, one can calculate arbitrary (even order) polynomials
in $\mathfrak{w}$ and $\mathfrak{w}^*$ in terms of the above
tensor operators. In particular, taking the sum of these two
relations, we get the following anticommutator for the baryonic
observables:
\begin{equation}
\label{comm-w*w}
\left[ {\mathfrak{w}}^*{}^{ijk} \,, \,  \mathfrak{w}_{pqr} \right]_+ =
\frac{1}{4} \left( Y^{i j k}{}_{p q r} - Z^{i j k}{}_{p q r} +
\tfrac{2}{3} D^{i j k}{}_{p q r} \right) \, .
\end{equation}
In fact, these relations (\ref{comm-w*w}) together with (\ref{commrel-j}), (\ref{commrel-jw}), 
(\ref{commrel-jw*}) and the mutual anticommutativity of $\mathfrak{w}$ and $\mathfrak{w*}$
can again be taken as the defining relations for a type of polynomial superalgebra generalising gl(12N/1),
this time with odd generators of antisymmetric type 
(see \cite{JR} for details, and also the discussion above).


\section*{Acknowledgments}

This research was partly supported by the Polish Ministry of
Scientific Research and Information Technology under grant
PBZ-MIN-008/P03/2003. J.~K. is grateful to
Professor E.~Zeidler for his hospitality at the Max Planck
Institute for Mathematics in the Sciences, Leipzig, Germany. 
G.~R. is grateful to the Foundation for Polish
Science (FNP) and to the Zygmunt Zaleski Foundation for their
support. He also wishes to thank the School of Mathematics and Physics at the  
University of Tasmania for its warm hospitality, and the Alexander von Humboldt Foundation for travel support. P.~D.~J. acknowledges the Australian Research Council, grant DP0208808 for 
support during the course of this work, and also thanks the Alexander von Humboldt Foundation for supporting visits to the Institute for Theoretical Physics, Leipzig. P.~D.~J. warmly thanks the Institute for Theoretical Physics, Leipzig and also the Max Planck Institute for Mathematics in the Sciences, Leipzig for hospitality and financial support during visits.

\end{document}